\documentclass[acmtog]{acmart}
\usepackage{arydshln}
\usepackage{multirow}
\usepackage{threeparttable}
\usepackage[percent]{overpic}
\usepackage{arydshln}
\usepackage{booktabs}
\usepackage{amsmath}
\usepackage{xspace}
\usepackage{wrapfig}
\usepackage[ruled]{algorithm2e}

\usepackage{bm}
\usepackage{makecell}
\usepackage{listings}
\usepackage{xcolor}
\usepackage{multirow}
\usepackage{float}
\usepackage{caption}
\usepackage{subcaption}

\usepackage{ulem}

\AtBeginDocument{%
  \providecommand\BibTeX{{%
    \normalfont B\kern-0.5em{\scshape i\kern-0.25em b}\kern-0.8em\TeX}}}

\setcopyright{acmcopyright}

\setcopyright{acmlicensed}
\acmJournal{TOG}
\acmYear{2024} \acmVolume{43} \acmNumber{6} \acmArticle{242} \acmMonth{12}\acmDOI{10.1145/3687904}

\newcommand{\name}{{CBIL}\xspace}

\begin{document}

\title{CBIL: Collective Behavior Imitation Learning for Fish from Real Videos}

\renewcommand{\shortauthors}{Wu*, Dou* et al.}

\newcommand{\TK}[1]{{\color{red}TK:#1}}
\newcommand{\ZY}[1]{{\color{red}ZY:#1}}
\newcommand{\ZYQ}[1]{{\color{red}ZY-\textbf{Q}:#1}}
\newcommand{\YK}[1]{{\color{gray}YK:#1}}
\newcommand{\YM}[1]{{\color{red}check:#1}}
\newcommand{\YF}[1]{{\color{orange}YF:#1}}
\newcommand{\TODO}[1]{{\color{red}TODO:#1}}
\newcommand{\todo}[1]{{\color{red}todo:#1}}
\newcommand{\update}[1]{{\color{blue}#1}}
\newcommand{\YFrep}[2]{\textcolor{orange}{\sout{#1}#2}}
\newcommand{\SO}[1]{{\color{magenta}{SO-\textbf{edit}:#1}}}
\newcommand{\SB}[1]{{\color{magenta}{SB-\textbf{edit}:#1}}}

\newcommand{\mba}{\mathbf{a}}
\newcommand{\mbs}{\mathbf{s}}
\newcommand{\mbq}{\mathbf{q}}
\newcommand{\mbp}{\mathbf{p}}
\newcommand{\mbv}{\mathbf{v}}
\newcommand{\mbg}{\mathbf{g}}
\newcommand{\mbd}{\mathbf{d}}
\newcommand{\mbz}{\mathbf{z}}

\author{Yifan Wu}
\authornote{\ \  Equal contribution.}
\affiliation{   \institution{The University of Hong Kong}
\country{Hong Kong}
}\email{wuyifan1@hku.hk}
\author{Zhiyang Dou}
\authornotemark[1]
\affiliation{  \institution{The University of Hong Kong}
\country{Hong Kong;}
\institution{University of Pennsylvania}
\country{U.S.A.}
}
\email{frankzydou@gmail.com}

\author{Yuko Ishiwaka}
\affiliation{  \institution{SoftBank Corp.}
\country{Japan}}\email{yuko.ishiwaka@g.softbank.co.jp}

\author{Shun Ogawa}
\affiliation{  \institution{SoftBank Corp.}
\country{Japan}}\email{shun.ogawa01@g.softbank.co.jp}

\author{Yuke Lou}
\affiliation{\institution{The University of Hong Kong}
\country{Hong Kong}}\email{louyuke@connect.hku.hk}

\author{Wenping Wang}
\affiliation{  \institution{Texas A\&M University}
\country{U.S.A.}}\email{wenping@tamu.edu}

\author{Lingjie Liu}
\affiliation{  \institution{University of Pennsylvania}
\country{U.S.A.}}\email{lingjie.liu@seas.upenn.edu}

\author{Taku Komura} 
\affiliation{  \institution{The University of Hong Kong}
\country{Hong Kong}}
\email{taku@cs.hku.hk}



\begin{teaserfigure}
  \includegraphics[width=\textwidth]{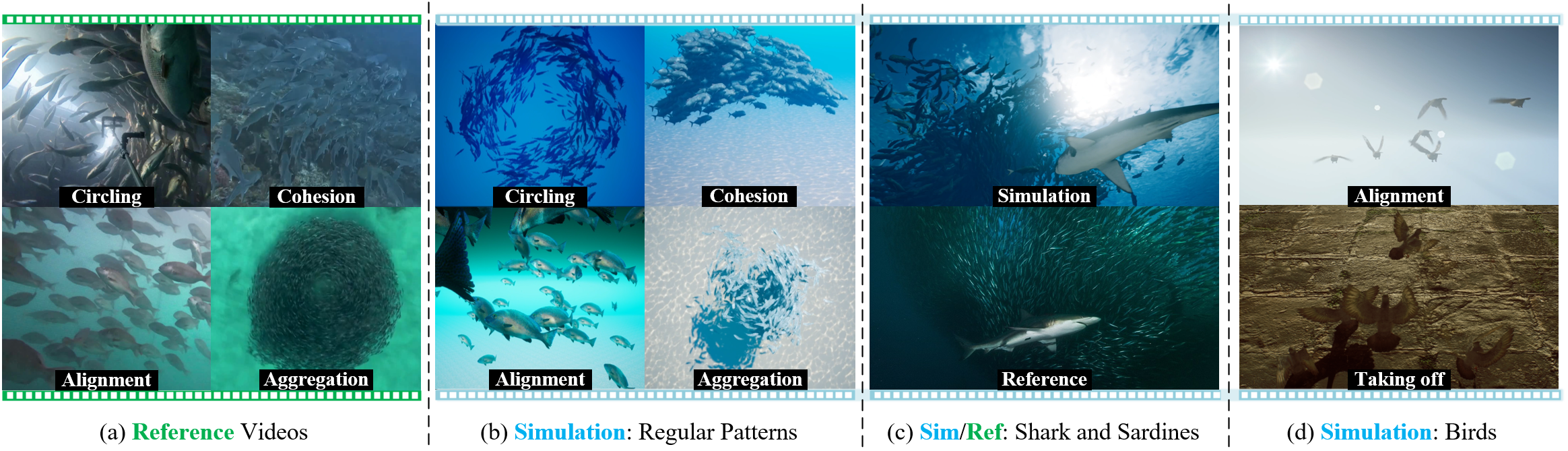}
  \vspace{-4.5mm}
  \caption{\name learns diverse collective behaviors of simulated fish from video inputs directly, enabling real-time synthesis of diverse collective motions. (a) reference video clips; (b) simulating varied behaviors of fish schools such as circling, alignment, cohesion, and aggregation; (c) fish schools responding to external changes and interactions; (d) motion control across different species, e.g., birds.}
  \Description{Teaser Figure.}
  \label{fig:teaser}
\end{teaserfigure}

\begin{abstract}
Reproducing realistic collective behaviors presents a captivating yet formidable challenge. Traditional rule-based methods rely on hand-crafted principles, limiting motion diversity and realism in generated collective behaviors. Recent imitation learning methods learn from data but often require ground-truth motion trajectories and struggle with authenticity, especially in high-density groups with erratic movements. In this paper, we present a scalable approach, Collective Behavior Imitation Learning~(CBIL), for learning fish schooling behavior \textit{directly from videos}, without relying on captured motion trajectories. Our method first leverages Video Representation Learning, in which a Masked Video AutoEncoder~(MVAE) extracts implicit states from video inputs in a self-supervised manner. The MVAE effectively maps 2D observations to implicit states that are compact and expressive for following the imitation learning stage. Then, we propose a novel adversarial imitation learning method to effectively capture complex movements of the schools of fish, enabling efficient imitation of the distribution of motion patterns measured in the latent space. It also incorporates bio-inspired rewards alongside priors to regularize and stabilize training. Once trained, \name can be used for various animation tasks with the learned collective motion priors. We further show its effectiveness across different species. Finally, we demonstrate the application of our system in detecting abnormal fish behavior from in-the-wild videos.
  \end{abstract}

\maketitle
\section{Introduction}
\label{sec:intro}
Reproducing realistic behaviors of fish schools offers a fascinating glimpse into the intricacies of collective behaviors observed in nature. The research not only deepens our understanding of the underlying principles governing the coordinated movements of fish~\cite{doi:10.1073/pnas.1109355108, Couzin2005, doi:10.1073/pnas.2320239121, CAVAGNA20181,couzin2002collective, ballerini2008interaction, cavagna2010scale,  newbolt2019flow, verma2018efficient} but also holds significant implications for various fields, such as robotics~\cite{zhou2022swarm, chung2018survey, kushleyev2013towards}, animation~\cite{ki2024learning,getz2024information}, as well as ecology and environmental science~\cite{liu2022close, guo2023student, zhang2024analysis, dell2014automated, hofmann2014evolutionary}.

\begin{wrapfigure}{r}{3.48cm}
\vspace{-4.3mm}
  \hspace*{-3mm}
  \centerline{ \includegraphics[width=37.3mm]{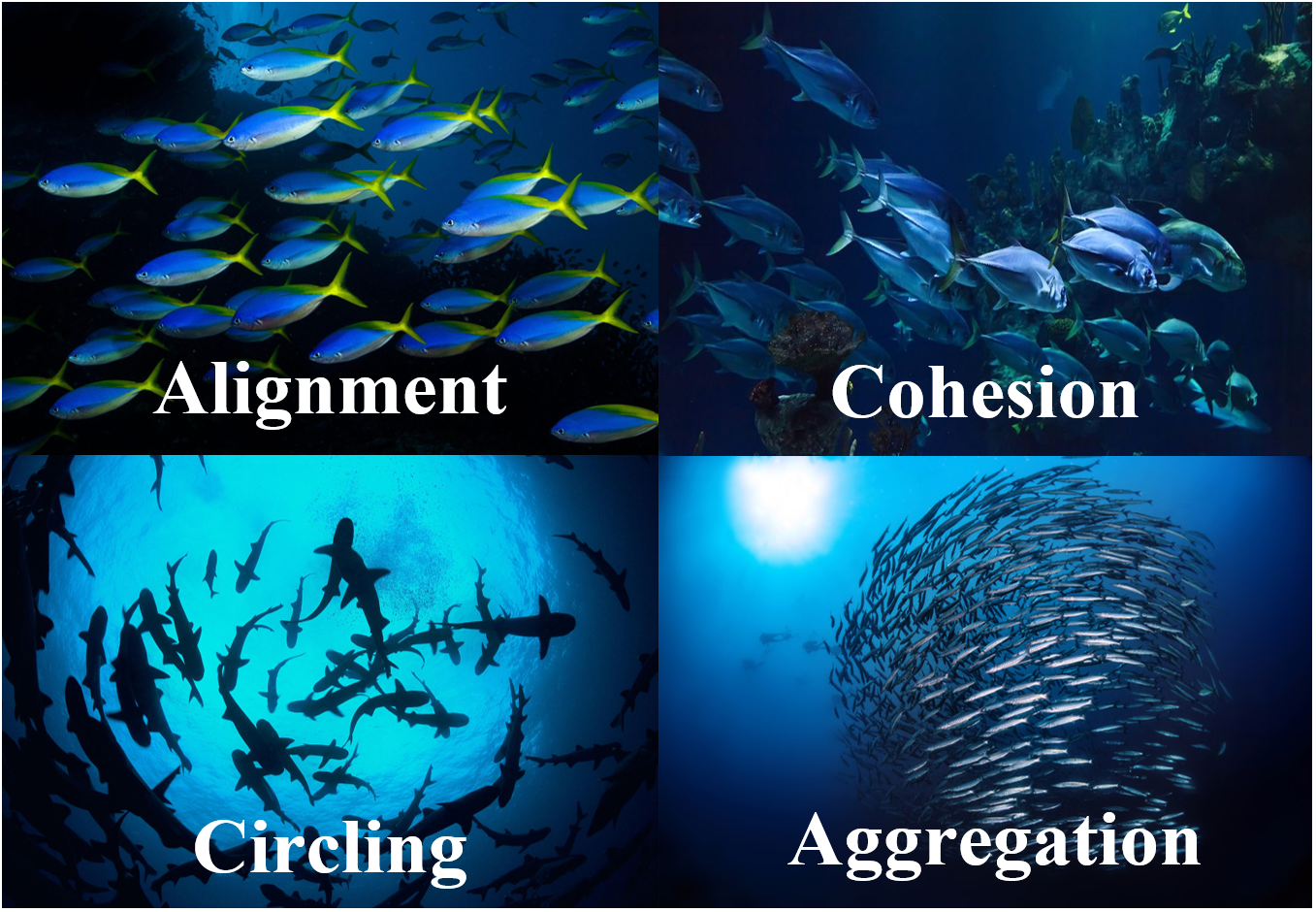}}
  \vspace{-3.5mm}
  \caption{\label{fig:diversefish} Diverse Fish Behaviors.}
  \vspace*{-5mm}
\end{wrapfigure}
Previous studies on simulating collective behaviors have been evolving over decades. Specifically, Boids \cite{10.1145/37402.37406} simulates flocking behavior using three hand-crafted rules. Foids \cite{10.1145/3478513.3480520} proposes a bio-inspired method, incorporating more physical information, e.g., boundary constraint, lighting, and temperatures, to simulate the movement patterns of fish in diverse environments. As a follow-up, DeepFoids~\cite{NEURIPS2022_74fa9e6b} further structures the behavioral model of fish as a rule-based crowd simulation by using Deep Reinforcement Learning. Overall, the aforementioned rule-based approaches have shown promising results in animating schools of fish. 
That being said, these hand-crafted rules still struggle to capture the intrinsic moving patterns due to the highly diverse, complex, and stochastic nature of fish school movements~(see Fig.\ref{fig:diversefish}). The diversity and randomness in their motion patterns make it a cumbersome task to reproduce the movement with high fidelity, especially when simulating with pre-defined rules, which further constrained their applicability in real-world scenarios. 

\indent Different from the rule-based method, data-driven approaches could capture the movement for reproducing diverse motions from real-world data. Data-driven crowd animation has been widely studied to reproduce diverse collective behaviors for fish~\cite{calovi2015collective}, butterflies~\cite{li2015biologically}, birds~\cite{doi:10.1073/pnas.1118633109} and human crowds~\cite{10.1145/3274247.3274510,10.1145/3592459,gupta2018social, 10.5555/1272690.1272706, ji2024text}.
Nevertheless, these methods are limited by their reliance on ground-truth trajectories in 3D or 2D, which significantly constrains their performance in scenarios where precise motion state information is unavailable. For example, in fish schooling, capturing the motion trajectory of each fish poses a significant challenge due to severe occlusions and highly similar textures. Occlusion hinders existing tracking methods like YOLOv9~\cite{wang2024yolov9}, leading to inconsistent trajectory data. Thus, the scarcity of data and the noise introduced during capture limit the effectiveness of the aforementioned techniques.

In this paper, we develop a scalable framework named Collective Behavior Imitation Learning (CBIL) to learn diverse fish schooling behaviors within a simulation environment. In contrast to previous data-driven methods, \name learns the collective behaviors of fish directly from 2D in-the-wild videos \textit{without} the reliance on the 3D motion trajectories. To achieve this, we first introduce a Video Representation Learning scheme to learn the motion states directly from the reference video clips. Specifically, a Masked Video AutoEncoder (MVAE) is trained to extract low-dimensional latent features in low-dimensional latent space from video inputs in a \textit{self-supervised} manner based on temporal vision transformers (ViT). This approach enables us to obtain compact and expressive \textit{implicit states} from the videos for imitation learning. During the imitation learning of \name, a policy learns to control simulated agents (e.g., fish) using the implicit features as input for the generative adversarial imitation learning~(GAIL)~\cite{GAIL}. This differs from conventional GAIL approaches~\cite{Peng_2021,Peng_2022,dou2023c} that typically rely on high-quality reference motions. To tackle the problem of mode collapse~\cite{Peng_2022,GAIL} faced by adversarial imitation learning framework, which hinders the capture of complex intrinsic collective motion skills and styles\footnote{In this paper, skill or style refers to different schools of fish moving patterns.}, 
the reference implicit states are clustered into distinct groups in an unsupervised manner for adaptively adjusting discrimination reward weights in imitation learning. More specifically, implicit states observed more frequently in the reference video are assigned larger reward weights to enhance distribution matching, thereby encouraging the model to capture discriminative movement features and improve robustness against noise. In addition to data-driven rewards from videos, we incorporate a biologically-inspired rule-based reward~\cite{NEURIPS2022_74fa9e6b} to regularize and stabilize the training process.

Our framework learns collective motion priors from various 2D videos, enabling the synthesis of various schooling behaviors such as circling, alignment, aggregation, feeding, and chasing. We further demonstrate its versatility by applying it to different species, such as birds (see Fig.~\ref{fig:teaser}). We also showcase the application in detecting abnormal fish behaviors in real-world videos. In summary, our contributions are threefold:

\begin{enumerate}
    \item We introduce Collective Behavior Imitation Learning~(\name), a scalable approach that learns collective motion priors of fish schools \textit{directly} from videos, without relying on 3D crowd trajectory motion capture.
    \item We develop a video representation learning model, Masked Video AutoEncoder, to facilitate adversarial imitation learning by capturing compact and expressive implicit states in a self-supervised manner.
    \item We present a method to efficiently capture the motion distribution of different crowd movement styles through implicitly latent clustering during the collective behavior imitation learning stage.
\end{enumerate}
\section{Related Work}
\paragraph{Collective Behavior Simulation}
Collective behavior simulation plays a crucial role in character animation and computer graphics, given its wide applications in character animation~\cite{rivers07, gustafson2016mure, headstrong12}, collective behavior simulation for animals, especially for fish~\cite{10.1145/37402.37406,10.1145/3478513.3480520, NEURIPS2022_74fa9e6b, MENG201855,10.1145/3023368.3036845, PhysRevLett.120.198101, NIWA199647,vicsek2012collective,aoki1982simulation}. It has also been a focus in analyzing collective behaviors in biological organisms~\cite{liu2022close, guo2023student, zhang2024analysis, dell2014automated, hofmann2014evolutionary, doi:10.1073/pnas.1109355108,10.7554/eLife.12852,PhysRevE.107.024411, Couzin2005, doi:10.1073/pnas.2320239121,CAVAGNA20181,couzin2002collective, ballerini2008interaction, cavagna2010scale}.

For human crowd animation, Lee et al.~\shortcite{10.1145/3274247.3274510} achieve crowd navigation using agent-based deep reinforcement learning. Leveraging deep neural networks such as convolutional neural networks, they navigate agents in dynamic environments with a single unified policy and a simple reward function, thereby eliminating the need for scenario-specific parameter tuning. Charalambous et al.~\shortcite{10.1145/3592459} learn a model for pedestrian behaviors guided by reference crowd data, obtaining a distribution of states extracted from real crowd data. 

For collective behavior simulation of animals, the seminal work Boids~\cite{10.1145/37402.37406} models bird flocks using three simple rules for spatial coordination and interaction, showing impressive results. Based on similar modeling ideas, collective motion of fish schools is also studied for scientific interest~\cite{aoki1982simulation,NIWA199647, PhysRevLett.120.198101,vicsek2012collective}. This approach lays the foundation for further research in simulating collective animal behaviors. Building upon this work, \cite{10.1145/3023368.3036845} extends the model by introducing predator-prey relationships, thereby enhancing motion diversity within the simulated population. Additionally, Satoi et al.~\shortcite{10.1145/2897824.2925977} propose a trajectory planning method incorporating a tube for Boids simulation, providing artists with more control over the animation process. Expanding upon these methods, Ishiwaka et al.~\shortcite{10.1145/3478513.3480520} utilize a rule-based approach to mimic biological motion patterns more accurately. Furthermore, Ishiwaka et al.~\shortcite{NEURIPS2022_74fa9e6b} present a method for synthesizing realistic underwater scenes with diverse fish species in various fish cages. They address the challenge of obtaining labeled datasets by introducing an adaptive bio-inspired fish simulation using Deep Reinforcement Learning. 

\paragraph{Imitation Learning for Physics-based Animation}
Imitation learning has shown its effectiveness in training agents to perform various tasks by observing demonstrations collected from experts. It has been extensively studied for physics-based character animation in the past decades~\cite{10.1145/3355089.3356536,10.1145/3478513.3480527,10.1145/3355089.3356501, lee2010data, liu2016guided,liu2010sampling,peng2018deepmimic, Peng_2021, Peng_2022, dou2023c, won2022physics, wang2023learning, lee2021learning, xu2023composite, park2019learning, won2020scalable, Tessler_2023, yao2022controlvae, feng2023musclevae, wang2024pacer+,pan2023synthesizing, yao2023moconvq}. Specifically, DeepMimic~\cite{peng2018deepmimic} trains simulated characters to acquire skills by mimicking reference motion clips. 
Adversarial motion priors in a GAIL style has been developed for body motion~\cite{Peng_2021, Peng_2022} as well as body-part level motion~\cite{bae2023pmp}. C$\cdot$ASE improves adversarial skill embedding efficiency in GAIL by learning a conditional skill distribution. AdaptNet~\cite{xu2023adaptnet} further enhances policy adaptation after imitation learning. 

However, most existing efforts in character controllers using imitation learning are for one single character or a relatively small group of people~\cite{zhang2023simulation,won2021control,rempe2023trace}. Imitation learning for crowd simulation has drawn researchers' attention~\cite{zou2018understanding,10.5555/1272690.1272706}. For instance, Zou et al.~\shortcite{zou2018understanding} propose a framework that involves a Recurrent Neural Network and trains a discriminator to learn plausible crowd-moving patterns from human trajectory data. As of yet, the aforementioned methods typically rely on relatively high-quality reference motions for imitation learning, utilizing systems such as motion capture~\cite{liu2016guided,peng2018deepmimic, Peng_2021, won2020scalable,won2021control,rempe2023trace}, pose estimation~\cite{cao2019openpose}, optical flow~\cite{HORN1981185}, trajectory detection~\cite{wang2024yolov9}, and synthesized body movements~\cite{zhou2023emdm, wan2023tlcontrol, guo2022generating, tevet2022human, zhang2024motiondiffuse} from generative models~\cite{luo2023perpetual,yuan2023physdiff,cong2024laserhuman, luo2023universal}. In contrast with these approaches, this paper presents the first GAIL-based framework for learning collective motion priors for large-scale swarms directly from video input.

\paragraph{Imitation Learning from Videos}
Previous research has delved into various methods for learning motion patterns from video clips. For instance, Vondrak et al.~\shortcite{10.1145/2185520.2185523} design a system tailored for physics-based character animation for video imitation. They utilize hand-crafted FSM controllers and an incremental optimization strategy focused on a 2D-silhouette matching objective. Later, Peng et al.~\shortcite{peng2018sfv} integrate 2D/3D pose estimators with deep reinforcement learning to train controllers capable of mimicking skill trajectories extracted from short video clips using~\cite{peng2018deepmimic}. Furthermore, Yu et al.~\shortcite{10.1145/3478513.3480504} extend this approach to replicate longer video sequences featuring dynamic camera movements and unpredictable environments. Additionally, Zhang et al.~\shortcite{zhang2023vid2player3d} introduce a hybrid control policy that refines learned motion embeddings by incorporating adjustments predicted by a higher-level policy, thereby enhancing the quality of motions extracted from broadcast videos through physics-based imitation. However, these methods primarily depend on pose estimation and tracking techniques, which may face challenges in collective behavior imitation scenarios due to the large number of fish with highly similar textures and significant occlusions. Inspired by recent advances in visual representation learning~\cite{he2022masked, tong2022videomae, caron2021emerging, oquab2023dinov2, he2020momentum}, which have made significant strides in capturing critical features from visual inputs, we propose a method to learn challenging collective motion priors \textit{directly} from videos for imitation learning, eliminating the need for traditional motion capture for the target collective behaviors.
\section{Collective Behavior Imitation Learning from Videos}
\label{sec:method}

\begin{figure}[t]
    \centering
    \begin{minipage}[t]{1\linewidth} 
        \centering
        \includegraphics[width=\linewidth]{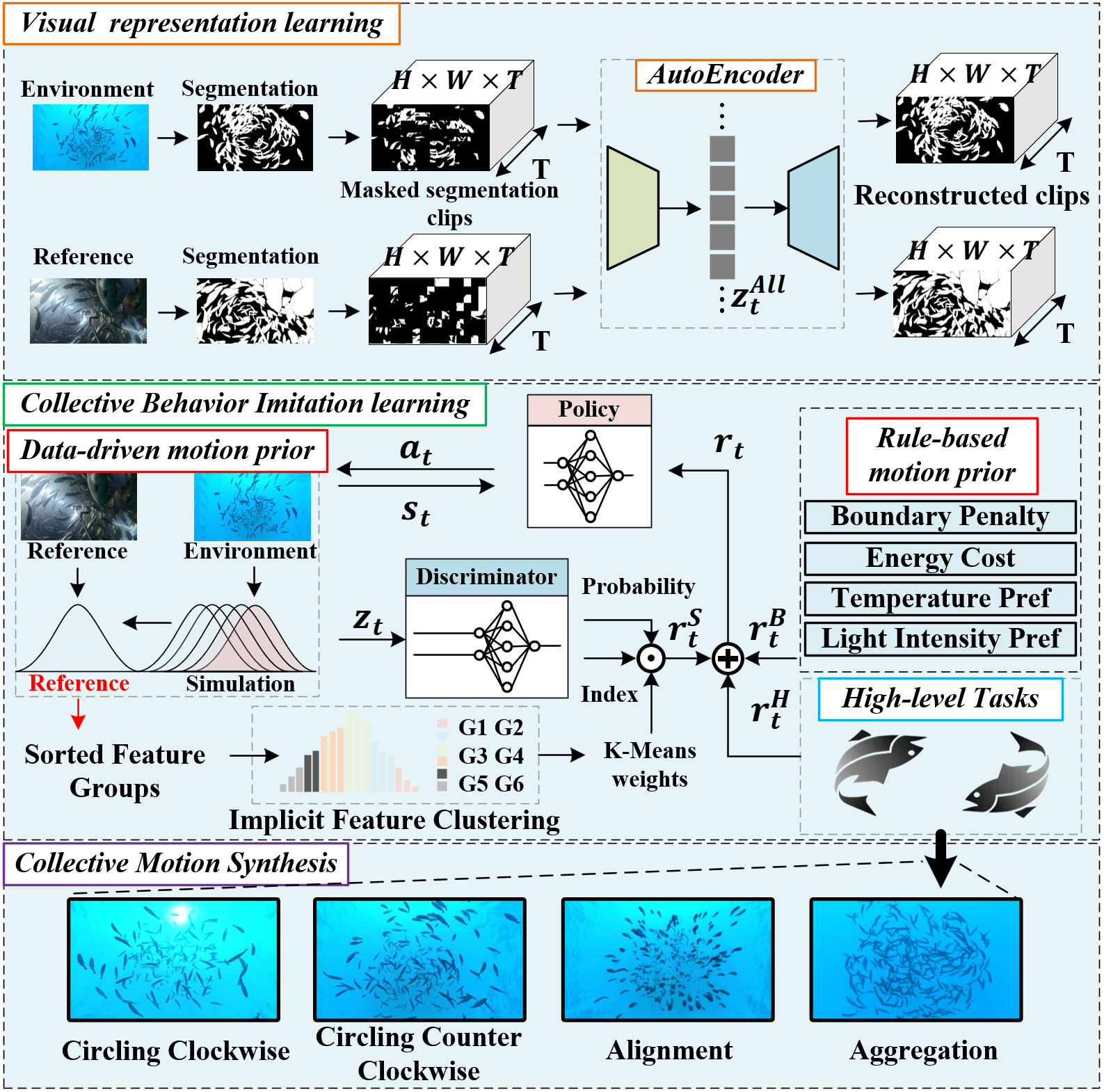}
                \vspace{-5mm}
        \caption{An overview of \name. Our framework has three stages: the visual representation learning stage, the collective behavior imitation learning stage, and the crowd animation stage for various animation tasks. In the first stage, we train the MVAE to learn mappings from video inputs to latent states. These latent states are later used in our collective behavior imitation learning. In the second stage, we employ both data-driven motion prior learned from videos and bio-inspired motion prior for imitation learning. Finally, we demonstrate that \name is applicable to diverse fish school animations such as circling, alignment, and aggregation.}
        \vspace{-2mm}
        \label{fig:Pipeline}
    \end{minipage}
\end{figure}

We introduce the Collective Behavior Imitation Learning~(\name) framework for simulated fish animation. An overview of our framework is shown in Fig. \ref{fig:Pipeline}, which includes the \textit{Visual Representation Learning} stage, the \textit{Collective Behavior Imitation Learning} stage, and the \textit{Collective Motion Synthesis} stage.

\textbf{\romannumeral1)} During visual representation learning, both reference videos and rendered results from the simulator are segmented and randomly masked first, producing masked segmented clips. We use a Masked Video AutoEncoder~(MVAE) to learn mappings between these video clips and implicit states of the crowd; See Sec.~\ref{sec:vrl}.

\textbf{\romannumeral2)} In the Collective Behavior Imitation Learning stage, our framework effectively captures diverse collective motion distributions by clustering the learned implicit states for adversarial imitation learning while integrating rule-based motion priors to regularize and stabilize the training process; See Sec.~\ref{sec:CIL}.

\textbf{\romannumeral3)} Finally, we show how to use \name for synthesizing different schooling behaviors; See Sec.~\ref{sec:high-level}.

\subsection{Visual Representation Learning from Videos}
\label{sec:vrl}

We introduce Visual Representation Learning to extract visual features from videos for subsequent adversarial imitation learning. We use both \textit{reference video clips}\footnote{The reference video clips are sourced from real fish farms and YouTube; detailed statistics are in Sec.~\ref{sub_sec:datasets}} and \textit{rendered video clips} of simulated fish schools (see  Appendix C for the setup) to train the system.

All video frames are first segmented into binary images using the SAM method~\cite{kirillov2023segment}. This approach excludes background and fish color information to enhance disentanglement, thereby facilitating the capture of discriminative features. For synthetic videos of fish, we project each shape into silhouettes for video segmentation within the simulator. This alleviates issues like occlusion, tracking failures, and the limited generalization of earlier tracking techniques
%

Inspired by MAE~\cite{he2022masked}, we mask the video frames for training a Masked Video AutoEncoder~(MVAE) to extract discriminative features self-supervisedly from video clips. Specifically, the input video clips $\mathcal{V}$ after segmentation are represented as $\mathcal{F}^{H \times W \times T}$, where $H$ and $W$ are the frame resolutions. $T=10$ denotes the window size, indicating a sequence of $10$ consecutive frames. We randomly mask 50\% of the patches from the resized segmented clips before sending them to the encoder. It learns to reconstruct the missing pixels in each segmented frame. Examples of video segmentation and masking are provided in Appendix F.

\begin{figure}[t]
    \centering
    \begin{minipage}[t]{1\linewidth} 
        \centering
        \includegraphics[width=\linewidth]{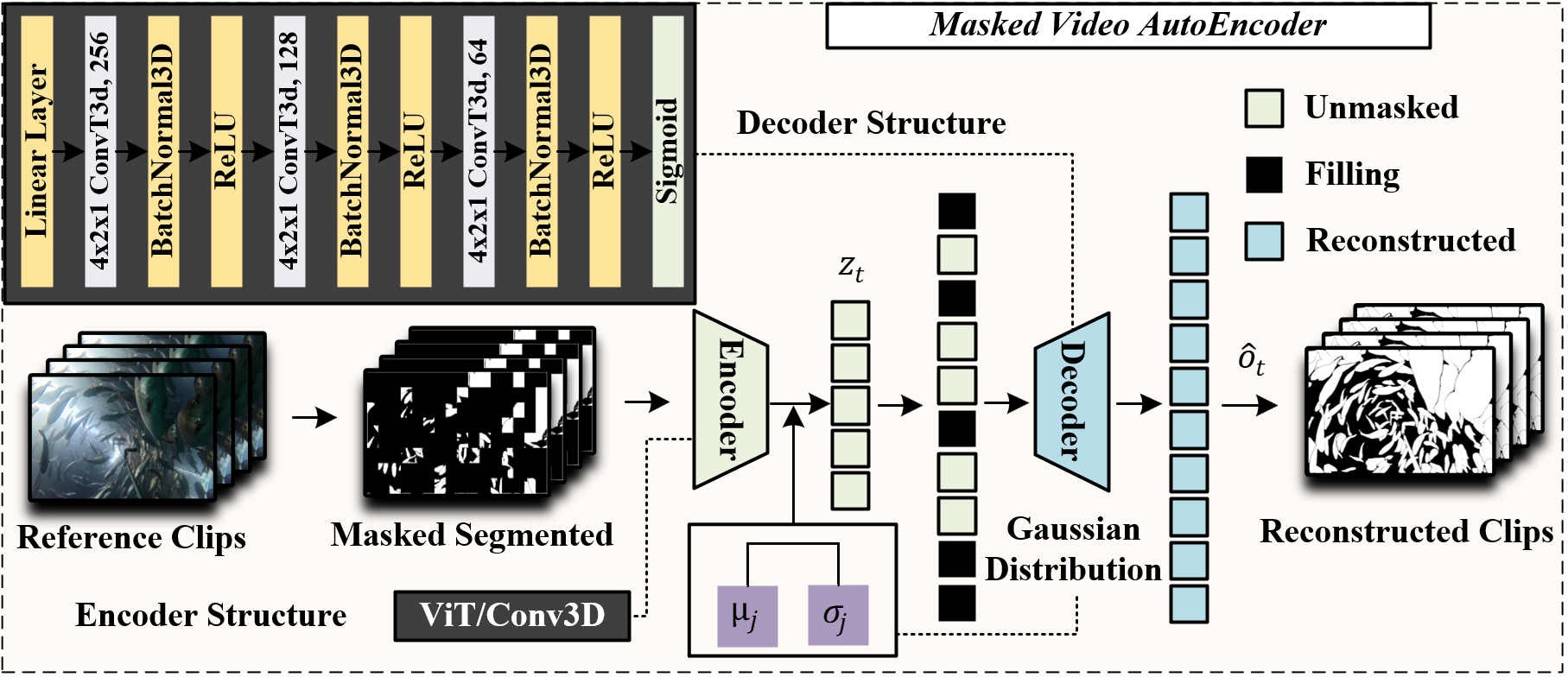}
                        \vspace{-5mm}
        \caption{Masked Video AutoEncoder. We use reference and rendered videos of simulated fish schools to train the model. Here, $z_t$ denotes low-dimensional implicit states with a dimensionality of 100, and $\hat{o}_t$ denotes reconstructed clips.}
        \vspace{-2mm}
        \label{fig:MVAE}
    \end{minipage}
\end{figure}

The network structure of the MVAE is shown in Fig.~\ref{fig:MVAE}. To obtain a compact and expressive manifold of the implicit states from video clips for adversarial imitation learning, we employ two loss functions: the reconstruction loss and the KL Divergence loss, as described below. 

\paragraph{Reconstruction Loss} The MVAE is trained to reconstruct video clips using the masked clips as input. The reconstruction loss is defined as follows:
\begin{equation}
    \mathcal{L}_R = \frac{1}{T}\sum_{t=1}^T\left(o_t-\hat{o}_t\right)^2,
\end{equation}

where $T=10$ is the window size of the video clips, $o_t \in \mathbb{R}^{H\times W\times C\times T}$ and $\hat{o}_t \in \mathbb{R}^{H\times W\times C\times T}$ are ground-truth clips and reconstructed clips respectively, and $C$ is the number of channels. We use the Mean Squared Error~(MSE) of $o_t$ and $\hat{o}_t$ as the reconstruction loss. 

The reconstruction loss ensures that the low-dimensional implicit states expressively represent the higher-dimensional features, i.e., video. We encourage the latent space to be compact using the KL Divergence loss to aid the discrimination during GAIL training.

\paragraph{KL Divergence Loss} 
KL divergence measures the difference between the latent space distribution in the VAE and a standard normal distribution:

\begin{equation}
    D_\textrm{KL}(Q(z|X) || P(z)) = \frac{1}{2} \sum_{j=1}^{J} \left(\mu_j^2 + \sigma_j^2 - \log(\sigma_j^2) - 1\right),
\end{equation}
where $X$ represents the input data, which is the input to the encoder network, \( \mu_j \) is the mean of the latent variable, \( \sigma_j \) is the standard deviation of the latent variable, and \( J \) is the dimensionality of the latent space, which is set to $100$ in this case.

\paragraph{Final Loss} The final loss for the autoencoder is defined as:
\begin{equation}
    \mathcal{L}_\textrm{Final} = \mathcal{L}_{R} + \beta  D_\textrm{KL}.
\end{equation}
This approach ensures that the MVAE learns an encoder capable of capturing discriminative motion characteristics with robust generalization. More implementation details of the MVAE training can be found in Appendix D.4.

\subsection{Collective Behavior Imitation Learning}
\label{sec:CIL}
With the MVAE, we introduce the Collective Behavior Imitation Learning framework for capturing the collective motion distribution from the video inputs.

\paragraph{Policy}
Our training framework is in a GAIL style, where the policy learns a collective motion prior from implicit states, which are produced by the pre-trained MVAE from the videos. The MVAE is frozen during this Collective Behavior Imitation Learning stage.
The policy $\pi$ learns to predict action and mimic the behaviors from demonstrations. The policy $\pi$ is used to control the individual fish but it is shared among all the fish. The policy is trained in an adversarial manner so that the discriminator cannot distinguish whether the behavior originates from the simulation or the reference. Specifically, the policy $\pi$ is given the fish's own state $\mbs_t$ as well as the states of its neighbors, and a goal $\mbg_t$, which is used as an observation/control signal to produce different patterns of collective behaviors as described in Sec.~\ref{sec:high-level}. 

The policy then outputs the action $\mba_t$ that follows a Gaussian distribution parameterized by the mean and covariance. The simulated agent, i.e., fish, then applies the action, which results in a new state $\mbs_{t+1}$ in the environment, as well as a scalar reward $r_t$~(Eq.~\ref{eq:rw_total}) which will be introduced in the following.

\paragraph{State Transition} 
Each fish agent has a state $\mbs_{t} \in \mathcal{S}$ at time step $t$ that consists of the forward direction $\mbd \in \mathbb{R}^3$, local position $\mbp \in \mathbb{R}^3$, rotation $\mathbf{q} \in \mathbb{R}^4$, and forward speed $v \in \mathbb{R}$. The goal is also passed to the policy for each task, which is detailed in Sec.~\ref{sec:high-level}. The action $\mba_t \in \mathcal{A}$, generated by the policies, transitions $\mbs_t$ to $\mbs_{t+1} \in \mathcal{S}$ through updating the forward velocity $\Delta v_t$ and rotation in the yaw and pitch axes, defined as $\Delta\theta^x_t$ and $\Delta\theta^y_t$. Additionally, the change in velocity, \(\Delta v_t\), is constrained within a range of allowable delta speeds, \(\hat \Delta v \in [0.8, 1.5]\) m/s, to maintain realism within the cage environment. Throughout this process, collisions between the fish agents and the cage boundary are also simulated.

\paragraph{Discriminator}
The discriminator is trained to effectively enforce the policy (generator) to reproduce reference behaviors in the simulator. Instead of using explicit 3D reference motion trajectories, as done in~\cite{Peng_2021,Peng_2022,dou2023c}, we train our discriminator using implicit state transitions $\mathcal{D}(\mathbf{z},\mathbf{z}')$, where the implicit states are extracted from video clips using MVAE. The discriminator is trained with the following objective:
\begin{equation}
    \min_{\mathcal{D}}  -\mathbb{E}_{d^{\mathcal{M}}(\mathbf{z},\mathbf{z}')} [\log \mathcal{D}(\mathbf{z},\mathbf{z}')] \\
 - \mathbb{E}_{d^{\mathcal{\pi}}(\mathbf{z},\mathbf{z}')} [\log(1 - \mathcal{D}(\mathbf{z},\mathbf{z}'))].     
\end{equation}
To improve robustness and effectiveness in adversarial imitation learning, which is often plagued by instability, we use gradient penalty regularization techniques inspired by the work of \cite{Peng_2021}. The discriminator is trained based on the following objective function:

\begin{equation}
\begin{split}
\min_{\mathcal{D}} 
&-\mathbb{E}_{d^{\mathcal{M}}(\mathbf{z},\mathbf{z}')} [\log \mathcal{D}(\mathbf{z},\mathbf{z}')] \\
&- \mathbb{E}_{d^{\mathcal{\pi}}(\mathbf{z},\mathbf{z}')} [\log(1 - \mathcal{D}(\mathbf{z},\mathbf{z}'))] 
\\
&+ w_{\text{gp}} \mathbb{E}_{d^{\mathcal{M}}(\mathbf{z},\mathbf{z}')} 
  \left[\left\| \nabla_{\varphi} \mathcal{D}(\varphi)|_{\varphi=(\mathbf{z},\mathbf{z}')} \right\|_2^2\right],
\end{split}
\end{equation}
where $w_\text{gp}$ is a manually specified coefficient, and $d^{\mathcal{M}}(\mathbf{z},\mathbf{z}')$ and $d^{\mathcal{\pi}}(\mathbf{z},\mathbf{z}')$ denote implicit state transitions $(\mathbf{z},\mathbf{z}')$ from reference skills and ones produced by the policy $\pi$, respectively. The policy is trained using the scaled probability of the discriminator $\mathcal{D}(\mathbf{z},\mathbf{z}')$ as the imitation reward.

\paragraph{Implicit State Clustering}
The GAIL framework \cite{Peng_2021,Peng_2022}, which includes a discriminator trained to discern whether a set of motions originates from the distribution of references, has suffered from mode collapse and has been unable to capture the frequency or entropy of the reference distribution, leading to low skill learning efficiency, as revealed by~\cite{dou2023c, Peng_2022, Peng_2021}. Previous studies~\cite{yao2022controlvae, dou2023c} have proposed explicitly learning conditional skill distributions to encourage the skill learning process, but acquiring the necessary skill labels for conditioning often poses significant challenges.

To cope with this problem in an unsupervised fashion, we employ feature clustering using K-Means to categorize all \textit{reference implicit states} mapped from MVAE into $N$ groups according to the distances of these latent features after dimensionality reduction using t-SNE, which enforces the network to produce more discriminative implicit motion states of the motion patterns. For example, if a motion state is clustered into a group of motion states frequently observed within the reference video, it is identified as having discriminative motion features and is assigned a larger reward weight for adversarial imitation. This training strategy learns the crucial features mapped from the videos and improves the robustness of our method against noisy references. Specifically, the implicit state clustering can be described using the following equations:
\begin{equation}
    r^\textrm{S}(\mathbf{z_t},\mathbf{z_{t+1}}, i) = -\log\left(1-\frac{W^\textrm{FG}_{i}  \mathcal{D}(\mathbf{z_t},\mathbf{z_{t+1}})}{\sum_{i=1}^N W^\textrm{FG}_i}\right),
    \label{eq:r_style}
\end{equation}

\begin{equation}
    W^\textrm{FG}_i = \frac{N_s^i}{\sum_i N_s^{i}},
    \label{eq:weight_details}
\end{equation}
where \( W^\textrm{FG}_{i} \) denotes the weight of implicit state transition group \( i \), which is calculated as the proportion of the number of implicit states in each reference group $N_s^i$ relative to the total number of implicit states in the entire reference set $\sum_iN_s^i$, $\mathcal{D}(\mathbf{z_t},\mathbf{z_{t+1}})$ denotes the output of discriminator, and we scale $r^s$ to $(0, 1)$ using a mapping function $\frac{2}{1+\text{exp}(r^S)}$. To ensure the reference implicit state transition distribution groups are sorted properly, we further incorporate the Sum of Squared Errors~(SSE) check:
\begin{equation}
    \textrm{SSE} = \sum_{i=1}^{K} \sum_{\mathbf{z} \in C_i} \|\mathbf{z} - \boldsymbol{\mu}_i\|^2,
\end{equation}
where $K$ is the number of clusters, automatically selected based on the SSE within different test $K$ values ranging from 1 to 10, $C_i$ is the $i$-th cluster and $\boldsymbol{\mu}_i$ is the centroid of the cluster. 
The elbow method \cite{8549751} is used to determine the optimal $K$ for K-means clustering. Cluster frequencies and centers are then stored for later use in assigning weights. We validate the effectiveness of this training scheme in Sec.~\ref{sec:feature_clustering}. 

\paragraph{Biological Rule-Based Rewards} 
In addition to the style reward learned from the videos in Eq.~\ref{eq:r_style}, we incorporate a biological rule-based reward \( r^\textrm{B}(\mbs_t, \mba_t, \mbs_{t+1}) \) as a regularization term for the fish agents, following \cite{NEURIPS2022_74fa9e6b}, to help stabilize the training process. Details of the rule-based rewards are in Appendix D.2.

\subsection{Synthesizing Specific Collective Motion Patterns}
\label{sec:high-level}
We define behavior-specific rewards as $r^\textrm{H}(\mbs_t, \mba_t, \mbs_{t+1}, \mbg_t)$, where the set $H$ includes circling ($r^\textrm{cir}$), aggregation ($r^\textrm{agg}$), alignment ($r^\textrm{ali}$), chasing ($r^\textrm{dom}, r^\textrm{sub}$), feeding ($r^\textrm{feed}$), and cohesion ($r^\textrm{coh}$).
The reward for each specific pattern is defined below. Note that the policy also receives a goal ${\mbg}_t$ as the control signal, which varies between the pattern: the variable given as a goal is denoted with the $^*$ superscript.   

\paragraph{Circling} In this pattern, the fish school exhibits clockwise movement (counterclockwise circling can be generated by mirroring the direction). The policy reward is computed based on the target direction ${\mbd}_t^*$, which is derived from the cross product of the vertical vector in the world coordinate system and the vector from the fish agent’s position to the cage center, and the desired velocity $v^*_t \in [0.8, 1.5]$m/s:

\begin{equation}
	r^\textrm{cir}_t = w_\text{circle}^d {\mbd^*_t}\cdot \frac{\mbv_t}{\|\mbv_t\|} -w_\text{circle}^v (\|\mbv_t\| - v^*_t)^2,
    \label{eq:circling}
\end{equation}
where $w_\text{circle}^d$ and $w_\text{circle}^v$ are the weights for direction and velocity control, both set to 10 for this experiment. $\mathbf{v}_t$ is the velocity of a fish at time step $t$.

\paragraph{Alignment}In this pattern, the fish attempts to align its velocity to those of its neighbors by setting its velocity towards the average velocity of its neighbors.  
As such, the reward is computed by: 

\begin{equation}
    r^\textrm{ali}_t = w_{ali}\sum_{j=1}^{|\mathcal{N}_t|} \frac{180^\circ - \theta^j_t}{180^\circ} 
\end{equation}
where $\mathcal{N}_t$ is the set of neighboring fish (within $3$-meter
radius of the agent fish) at time $t$, and $\theta_t^j$ is the angle between the forward direction of the current fish and the $j$-th neighboring fish, given by \(\theta^j_t = \angle (\mbd^\text{current}_t, \mbd^{j, \text{neighbor}}_t) \in [0^\circ, 180^\circ]\). The goal of alignment here is the normalized average forward direction of neighboring fish: $\mbd^*_{t, norm} = \frac{\mbd^*_t}{\|\mbd^*_t\|}$, where $\mbd^*_t = \frac{1}{|\mathcal{N}_t|}\sum_{j=1}^{|\mathcal{N}_t|} \mbd^{j,\text{neighbor}}_t$. We set the alignment weights, denoted as $w_{\text{ali}}$, to 1 in this experiment.

\paragraph{Aggregation} In this pattern, the policy directs each fish agent toward the school’s center, denoted as 
$\mbp_{t}^*= \frac{1}{|\mathcal{N}_t|}\sum_{j=1}^{|\mathcal{N}_t|} \mbp_t^j $, where $\mbp^{j}_t$ is the position of the $j$-th fish agent at time step $t$. We consider the neighboring fish within a 5-meter radius of the agent fish. The reward is defined as follows: 
\begin{equation}
r^\textrm{agg}_t = -\frac{w_\textrm{agg} \|\mbp_{t}-\mbp_{t}^*\|}{1 + e^{-a( \|\mbp_{t}-\mbp_{t}^*\| - b)}},
\end{equation}
where $a$ and $b$ are hyperparameters that adjust the degree of aggregation, and $w_\textrm{agg}$ is the aggregation weight. The goal observation for the policy is $\mbp_{t}^*$.

\paragraph{Chasing}
In this pattern, the dominant fish chases a subordinate fish.  
Here, the direction between the dominant and subordinate fish is given as the goal vector:  
$\mbd^*_t = \frac{\mbp^{\text{sub}}_t - \mbp^{\text{dom}}_t}{\|\mbp_t^{\text{sub}} - \mbp_t^{\text{dom}}\|}$,
where $\mbp_t^{\text{sub}}$ and $\mbp_t^{\text{dom}}$ represent the positions of the nearest subordinate fish relative to the dominant one, and the positions of the dominant fish relative to the subordinate fish at time $t$, respectively. The reward for the dominant fish is then defined by: 
\begin{equation}
	r^{\text{dom}}_{t} = w^\text{dom} \mbd^*_t \cdot \mbv_t^\text{dom}.
\end{equation}
where the weight $w_\text{dom} \in \mathbb{R}$ is set to $8$.

The reward for the subordinate fish is similarly computed but to swim away from the dominant fish:  
\begin{equation}
    r^{\text{sub}}_{t} = w^{\text{sub}} \mbd^*_t \cdot \mbv_t^{\text{sub}},
\end{equation}
where the weight $w_t^\text{sub} \in \mathbb{R}$ is set to $1$. 

\paragraph{Cohesion}
Here, the fish agent is attracted to the average location of its surrounding fish~(within 3-meter radius of the agent fish) $\mbp^*_t = \frac{1}{|\mathcal{N}_t|}\sum_{j=1}^{|\mathcal{N}_t|}\mbp^j_t$ by the following reward:  
\begin{equation}
    r_t^\textrm{coh} = w_\textrm{coh}     \|\mbp_t-\mbp^*_t\|,
\end{equation}
where the cohesion weight $w_\textrm{coh} \in \mathbb{R}$ is set to $5$. $\mbp^*_t$ is the target position for the fish agent, while $\mbp_t$ denotes the position of the controlled fish agent at time~$t$.

\paragraph{Feeding}

The fish agent is trained to move toward food positions, where it receives a reward upon collision with an object tagged as food. The reward function for feeding at time step \( t \) can be expressed as:
\begin{equation}
r^\textrm{feed}_t = 
\begin{cases} 
R_\textrm{feed}, & \|\mbp_t - \mbp^f_t\| < \epsilon \\
0, & \text{otherwise.}
\end{cases}
\end{equation}
Here, \( R_\textrm{feed} = 10 \) denotes the reward value assigned when the fish successfully feeds. The variables \( \mathbf{p}_t \) and \( \mathbf{p}^f_t \) denote the position of the fish agent, the position of the closest food item with respect to the fish agent at time $t$. \( \epsilon =0.01m\) is a threshold for determining if the fish has reached the food position. The goal observation is the distance vector between the fish agent and the nearest food item, given by \( \mathbf{p}^f_t - \mathbf{p}_t \). During training, the food position is randomly generated within the whole cage.

\subsection{Total Reward and Network Structure}
\label{sec:model_rep}

\paragraph{Total Reward}
Finally, the total reward function $r^t$ for policy training is defined as:
\begin{equation}
r^t =W_\textrm{S} 
    r^\textrm{S}(\mathbf{z_t},\mathbf{z_{t+1}},i) +
     W_\textrm{B}  r^\textrm{B}(\mbs_t,\mba_t,\mbs_{t+1})  + W_\textrm{H}  r^\textrm{H}(\mbs_t,\mba_t,\mbs_{t+1},\mbg_t), 
     \label{eq:rw_total}
\end{equation}
where the coefficients $W_\textrm{S}$, $W_\textrm{B}$ and $W_\textrm{H}$ are set to $0.4$, $0.1$, and $0.5$, respectively. All rewards are scaled to the range of $[0,1]$ during the policy training. The style reward $r^\textrm{S}_t$ (Eq.~\ref{eq:r_style}) is employed to learn styles from reference videos by extracting the implicit states using the MVAE, while the bio-inspired rule-based one $r^\textrm{B}$ aid in stabilizing the training process and replicating patterns influenced by biological environments. The task reward $r^\textrm{H}$, as outlined in this section, is used for various fish animations.

\paragraph{Network Structure} 
The policy $\pi$ is modeled by a neural network that maps given states $\mbs_t$ and the goal $\mbg_t$ to a Gaussian distribution over actions $(\mba_t|\mbs_t, \mbg_t) = \mathcal{N}(\mu(\mbs_t, \mbg_t), \Sigma_{\pi})$, with an input-dependent mean $\mu(\mbs_t, \mbg_t)$ and a fixed diagonal covariance matrix $\Sigma_{\pi}$, structured as a fully connected network with $4$ hidden layers of $1024$, $1024$, $1024$, and $512$ units.
\section{Experiment Settings}
\label{sec:exp}

The implementation of this system is based on Unity Engine~\cite{juliani2020unity} and ML-Agents\footnote{\url{https://github.com/Unity-Technologies/ml-agents}} with Python servers. Details are provided below.

\subsection{Simulation Platform}
\label{sec:simulation}
\begin{figure}[htbp]
    \centering
    \vspace{-1.5mm}
    \begin{minipage}[t]{1\linewidth} 
        \centering
        \includegraphics[width=0.8\linewidth]{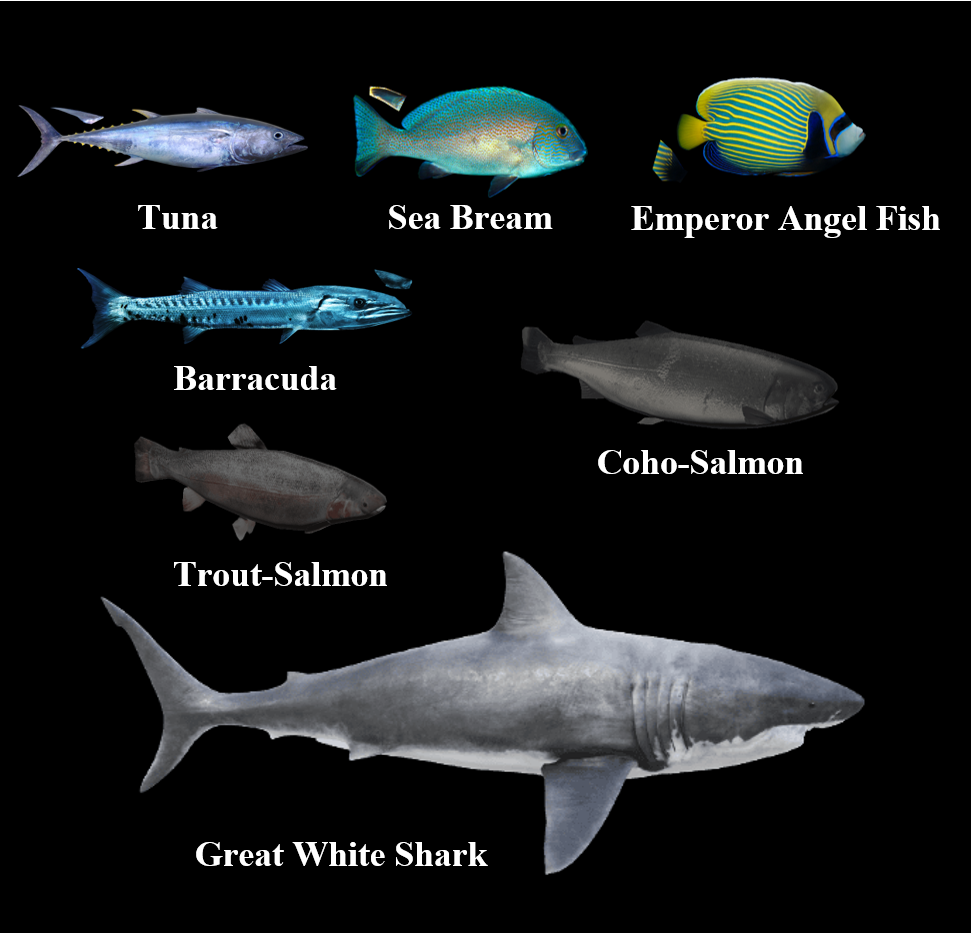}
                        \vspace{-2mm}
        \caption{Illustration of some of the 3D simulated fish models we utilize. Our scalable approach enables the training of policies for a broad spectrum of fish species.}
        \label{fig:fish_species}
    \end{minipage}
    \vspace{-4mm}
\end{figure}

In this paper, we simulate fish movement patterns with the Unity Engine~\cite{juliani2020unity}. Each fish agent is equipped with collision sensors to avoid neighboring fish, the ocean surface, and aquatic cages defined by volume boundary settings. Specifically, the physics simulation uses Unity components, such as UnityEngine.PhysicsModule and Unity Collider. We trigger the end of the episode whenever the fish agents collide with the cage boundaries or other fish agents. Note that only collision among fish agents and the cage boundary are simulated, while fluid dynamics and rigid body-fluid interactions are not accounted for in this simulation. In terms of biological realism, we account for fish species, size, quantity, and speed. The simulation runs on a laptop with an NVIDIA GeForce RTX 3070 Ti GPU, 64GB RAM, and a 12th Gen Intel Core i7-12700H processor on Windows 11. Fig.~\ref{fig:fish_species} illustrates some of the 3D simulated fish models we utilize. We provide a detailed breakdown of computation costs for the training and inference stages in Appendix A.

\subsection{Assets and Camera Setup for Animation}
\label{sib_sec:assets}
For visualization and rendering purposes, we use skinned fish characters as shown in Fig.~\ref{fig:fish_species}. 
The meshes and textures for the fish agents are manually crafted, e.g., the fish agent has 57 skeletal nodes for animation. Each fish is paired with an animation controller that produces detailed body movements, which are also manually designed. The predicted action from the policy automatically triggers the Unity animation controller to execute the predefined fish body part motion. Rendered results of the simulated fish school are generated by manually adjusting the camera parameters so that rendered results closely approximate the settings of real cameras. 
%


\subsection{Dataset} 
\label{sub_sec:datasets}
We evaluate our model using a self-collected dataset comprising underwater scenes captured by an Osmo Action 3 camera for the red sea breams in the fish cage whose size is 12m $\times$ 12m (width) $\times$ 9m (depth). The camera has a field of view (FOV) of 155° and a focal length of 12.7mm, with recordings spanning over a year. The dataset encompasses diverse weather and seasonal conditions and varying light intensities across different time periods and is captured from different perspectives with the same fish group.
To evaluate the robustness of the method, we also use datasets of other fish species, such as Trout-Salmon, Coho-Salmon, and Yellowtail, in fish cages, which are captured by KODAK 4KVR 360. Detailed information on our devices and settings are shown in Appendix D.1.2. The entire captured data spans approximately 2TB in size. Additionally, we utilize several YouTube videos to perform imitation tasks. 
The selected video clips, listed in Tab.~\ref{tab:dataset_list}, are divided into training and test datasets with a 4:1 ratio.

\begin{table}[ht]
  \caption{The reference video dataset used for model training.}
        \vspace{-3mm}  
  \label{tab:dataset_list}
  \footnotesize
\begin{tabular}{lcccc}
    \toprule
 Dataset & Frames & Motion Types \\
    \midrule
    Trout-Salmon& 590 & aggregation \\
    Coho-Salmon-1 & 726 & feeding, circling \\
    Coho-Salmon-2 & 629 & alignment, cohesion\\
    Shark and Sardines & 1208 & predation\\
    4k-Youtube & 5028 & aggregation, circling, alignment, cohesion\\
    Youtube-Birds & 1100 & alignment, walking\\
  \bottomrule
\end{tabular}
\end{table}

\subsection{Evaluation Metrics}
\label{sub_sec:metrics}
We use the following metrics to evaluate the performance of four collective motion synthesis applications: two circling patterns (clockwise and counterclockwise), alignment, and aggregation. Specifically, we evaluate using cross-view validation, skill distribution similarity, diversity analysis, and task return. Each metric is described below. We provide more detailed settings of metrics in Appendix D.6.

\paragraph{Cross-View Validation}
To examine our models' generalization ability on unseen views, we use Fréchet Inception Distance (FID) based on image features. Specifically, we train our models on reference videos with multiple views and test motion consistency using unseen views by comparing the generated frames with the ground-truth frames from the same reference dataset. The FID can be defined as:
\begin{equation}
    \text{FID} = \left\| \boldsymbol{\mu}_{\text{real}} - \boldsymbol{\mu}_{\text{gen}} \right\|_2^2 + \text{Tr}\left(\boldsymbol{\Sigma}_{\text{real}} + \boldsymbol{\Sigma}_{\text{gen}} - 2\left(\boldsymbol{\Sigma}_{\text{real}}\boldsymbol{\Sigma}_{\text{gen}}\right)^{1/2}\right), 
\end{equation}
where \( \boldsymbol{\mu}_{\text{real}}, \boldsymbol{\Sigma}_{\text{real}}  \) and \( \boldsymbol{\mu}_{\text{gen}},  \boldsymbol{\Sigma}_{\text{gen}} \) represent the mean and covariance matrices of the features of the real and generated frames, respectively, 
and \( \text{Tr}(\cdot) \) denotes the trace of a matrix. We use a pre-trained ResNet-50~\cite{he2015deep} to extract features from each image for FID computation, where each output feature has a dimension of $2048$.

\paragraph{Skill Distribution}
To examine models' ability to accurately learn from the reference collective motion distribution, we utilize Jensen-Shannon Divergence based on the video features which is extracted by MVAE (Sec.~\ref{sec:vrl}). It is the average of the KL divergences between two probability distributions and their average distribution, and less sensitive to outliers and small deviations between probabilities in the distributions. It is also bounded between 0 and 1, a value of 0 indicates that the two distributions are identical, while a value of 1 indicates maximum dissimilarity. 

\paragraph{Diversity Analysis}
We assess the diversity of the generated fish school animations. 
 Following~\cite{dou2023c,wang2022diverse,lu2022actionconditioned}, we utilize the Average Pairwise Distance (APD) to assess the diversity of a set of generated motion sequences, i.e., root trajectories. Specifically, given a set of generated motion clips from simulator $M = \{m_i\}$, each motion clip contains $l$ frames, APD is defined as 
\begin{equation}
    \textrm{APD}(M) = \frac{1}{N(N - 1)} \sum_{i=1}^{N} \sum_{j \neq i}^{N} \left( \sum_{t=1}^{l} \left\|\mbs_{i}^{t} - \mbs_{j}^{t}\right\|_{2} \right)^{1/2},
\end{equation}
where $\mbs_{i}^{t} \in M_{i}$ is a state as defined in Sec.~\ref{sec:model_rep} within a motion clip $M_{i}$ and $N$ is the number of generated sequences. A larger APD represents a more diverse set of motion clips.

\paragraph{Task Return}
Following ~\cite{peng2018deepmimic,Peng_2021}, we report task return, which represents the reward the agent receives during task execution. Task return is defined by the specific goals and constraints of the task, and a policy is trained to maximize the task return. In our experiments, we use normalized task returns to evaluate the performance of different methods.

\subsection{Training Details}
\label{sec:training_details}
In MVAE, the coefficient $\beta$ of KL Divergence loss is set to $0.5$. Further details of MVAE can be found in Appendix D. In CBIL, we use the Proximal Policy Optimization (PPO) algorithm \cite{schulman2017proximal} for training the policy during imitation learning. The actor and critic networks use network structures, each with $128$ hidden units per layer and consisting of 4 layers. The replay buffer size is set to $1 \times 10^6$. The learning rate is $3 \times 10^{-4}$ for tasks, $2 \times 10^{-4}$ for imitation and with a batch size of $1000$ for policy training. In PPO, we set the value of $\beta$ to $5 \times 10^4$, $\epsilon$ to 0.2, and the discount factor gamma ($\gamma$) to $0.99$. We apply Generalized Advantage Estimation (GAE) with parameter $\lambda=0.99$ for the estimation of the advantage function. The discriminator consists of an input layer with $128$ units, with $32$ hidden units, and an output layer with a single neuron. The network comprises two fully connected layers and utilizes the Tanh activation function to restrict the output to $[-1, 1]$. The Adam optimizer \cite{kingma2017adam} is employed for network training.

For training, we use $50$ fish agents, each initialized with random states, including position, rotation, and velocity within a limited range. An episode terminates when a fish triggers collision detection. The whole training process involves $4 \times 10^6$ simulation steps. Training is performed on a single GTX 3070Ti 8GB GPU, requiring approximately 10 hours to complete. The maximum number of simulation steps is set to $400,000$. 

During inference, users can manually specify the number of fish agents in the environment. Since our framework follows a Multi-Instance Single Policy scheme, the policy processes each fish's state individually during both training and inference. This ensures that the learned policy is adaptable to simulations with varying numbers of fish agents.

\section{Implicit States from Videos}
\label{sec:implicit_states}

\begin{figure}[htbp]
    \centering
    \begin{minipage}[t]{1\linewidth} 
        \centering
        \includegraphics[width=\linewidth]{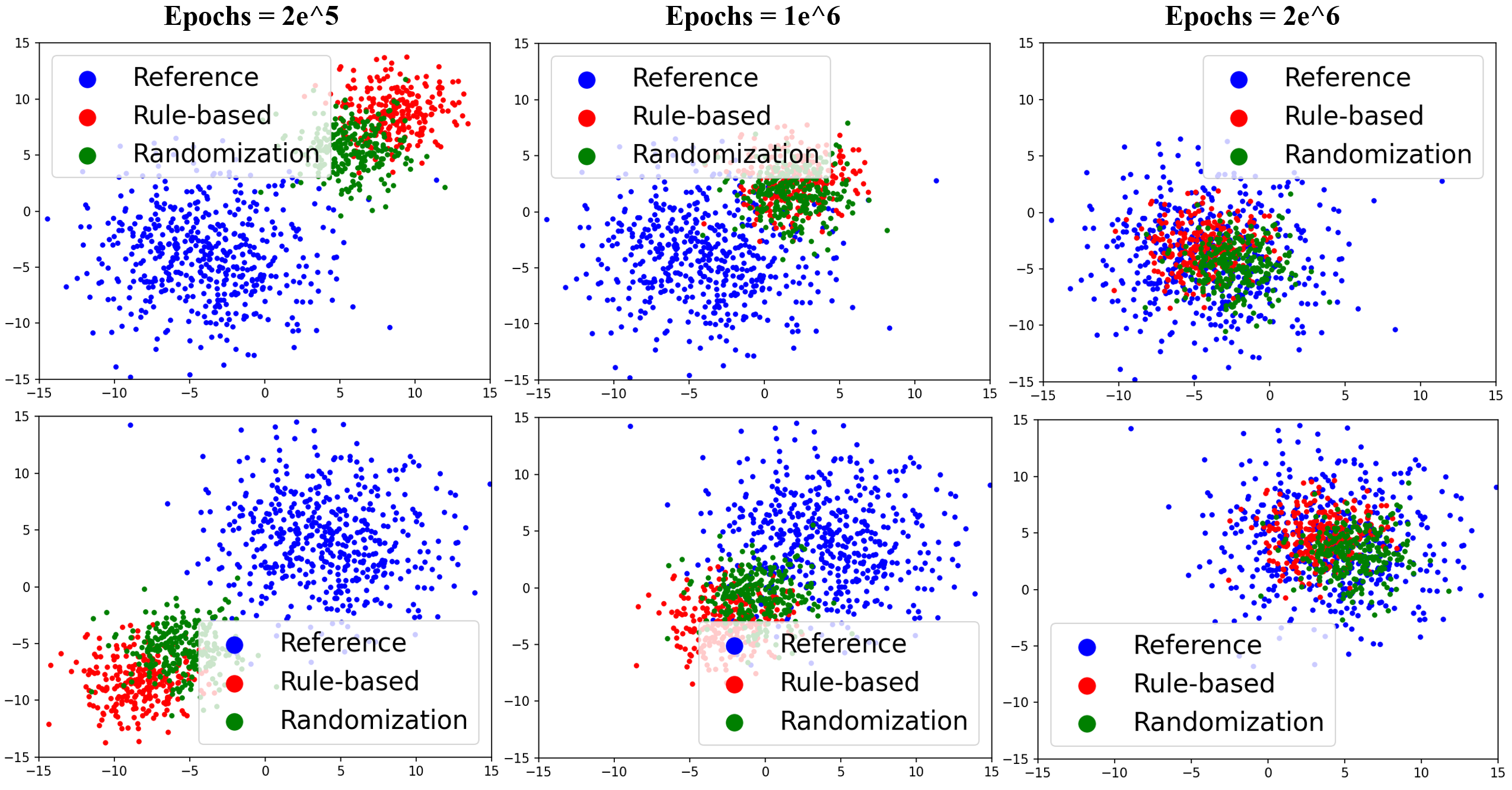}
        \vspace{-4mm}
        \caption{t-SNE visualization of MVAE latent representations for reference and simulated circling videos during pre-training. \textbf{Top}: Clockwise, \textbf{Bottom}: Counterclockwise. Colors represent different sources: \textcolor{blue}{blue} for reference, \textcolor{red}{red} for rule-based generated, and \textcolor{green}{green} for randomization.}
        \label{fig:tsne-latent}
    \end{minipage}
\end{figure}

The MVAE is trained on both reference and simulated videos to enhance its generalizability to various inputs. We investigate the generalization capability of the MVAE by visualizing the latent space computed from both the reference and simulation video clips by a t-SNE plot in Fig.~\ref{fig:tsne-latent}: the implicit state distribution from the simulator gradually converges and expands its coverage of implicit states across various video inputs throughout the MVAE training process. As a result, it sufficiently covers diverse crowd motion distributions from reference videos and rendered animations, enabling effective and scalable Collective Behavior Imitation Learning in the subsequent stage, where videos are mapped to compact latent spaces for discrimination a GAIL framework.

\section{Collective Motion Synthesis}
\label{sec:CMP}

\begin{figure*}[t]
    \centering
    \includegraphics[width=\linewidth]{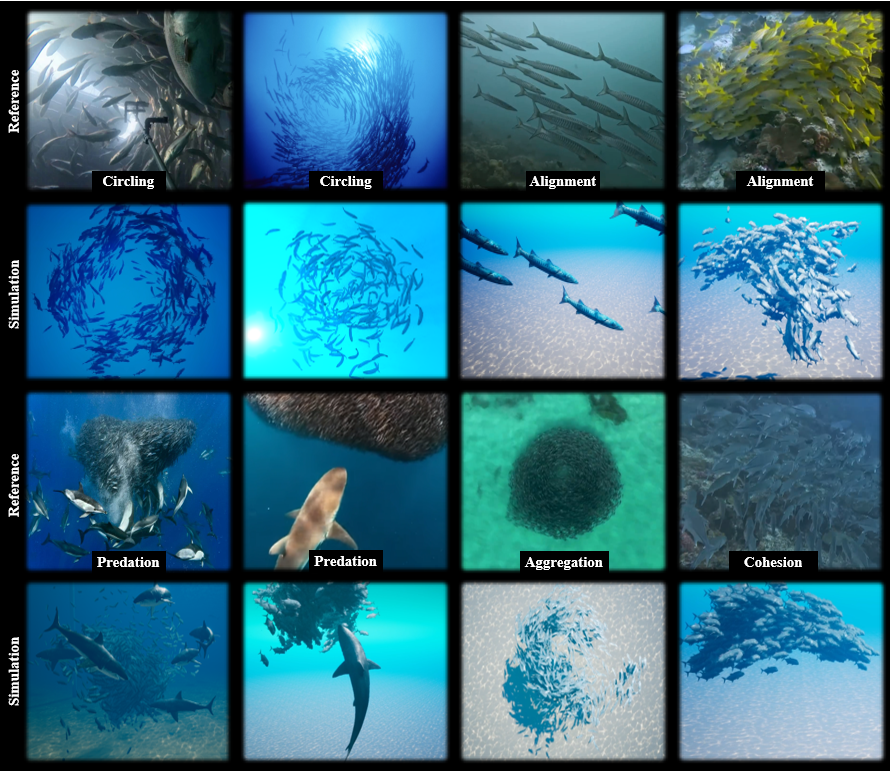}
    \vspace{-5mm}
    \caption{A gallery showcasing fish school animations reproduced by our method, highlighting its effectiveness in reproducing diverse behaviors across various fish species.}
    \label{fig:real2sim}
    \vspace{-3mm}
\end{figure*}

\begin{figure*}[t]
\vspace{-1mm}
    \centering
    \includegraphics[width=\linewidth]{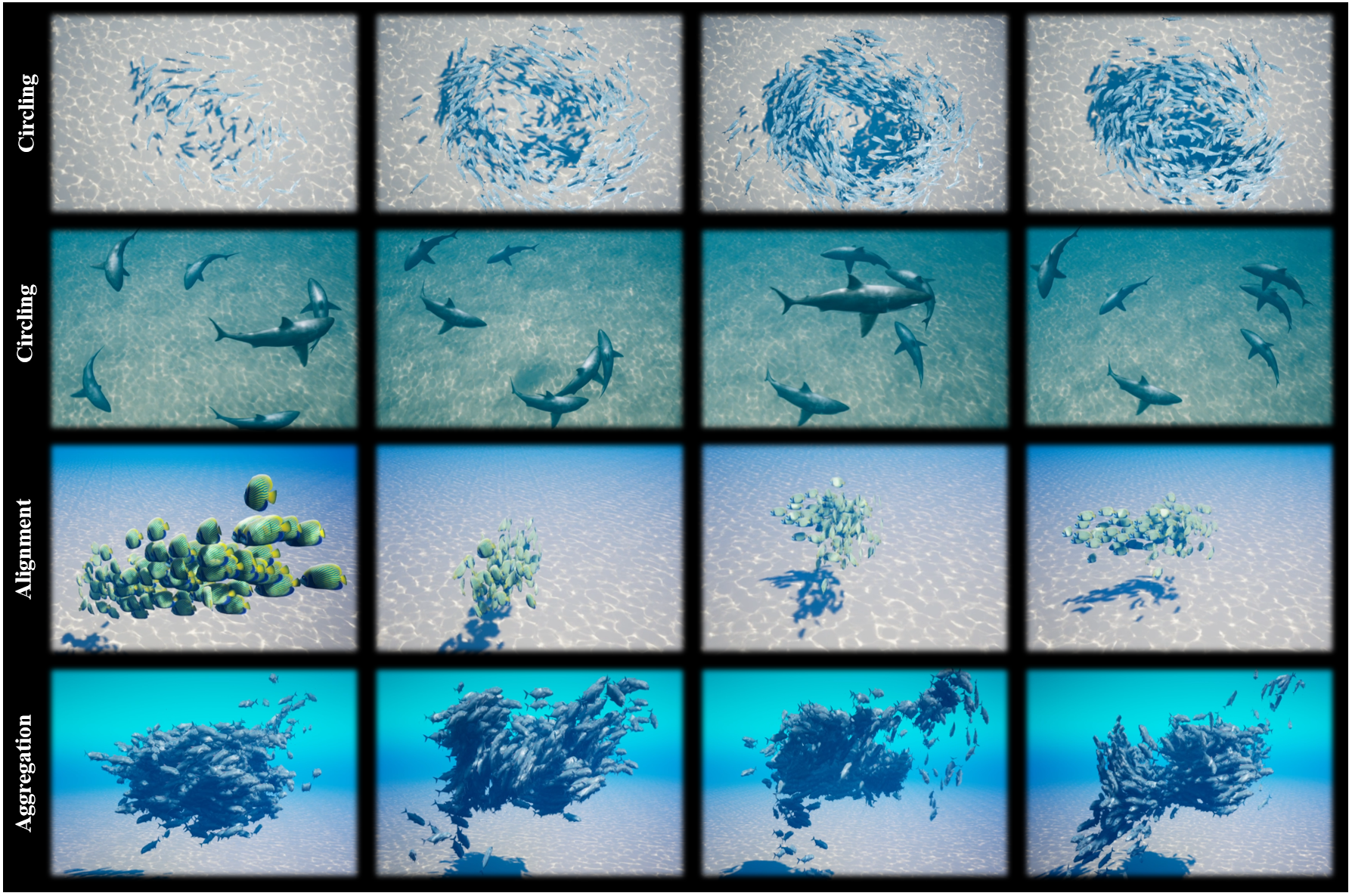}
    \vspace{-7mm}
    \caption{We showcase multiple animations of fish schools with various movement patterns: (a) Tuna circling clockwise with 50, 100, and 300 fish; (b) Shark circling counterclockwise; (c) Emperor Angel Fish alignment with 50 fish; (d) Sardines aggregation.}
    \label{fig:Result_highlevel}
    \vspace{-1mm}
\end{figure*}

\begin{figure}[t]
    \centering
    \begin{minipage}[t]{1\linewidth} 
        \centering
        \includegraphics[width=0.95\linewidth]{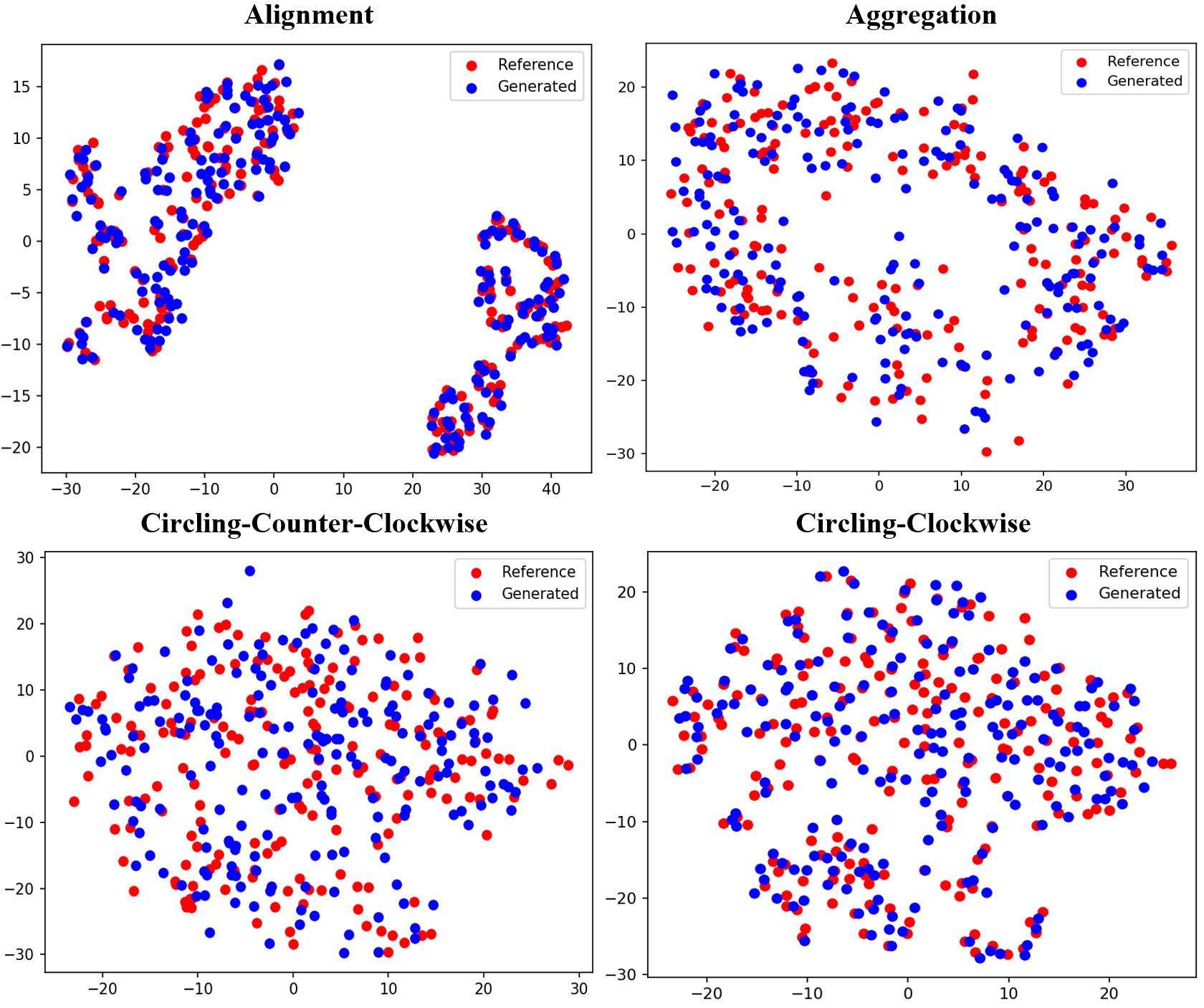}
                \vspace{-4mm}
        \caption{Qualitative result of our method: t-SNE of unseen \textcolor{red}{reference} video clips and \textcolor{blue}{generated} video clips of different tasks. Each reference and generated video takes 200 clips; the perplexity is set to 30.}
        \label{fig:FID}
    \end{minipage}
\end{figure}

\begin{figure}[t]
    \centering
    \begin{minipage}[t]{1\linewidth} 
        \centering
        \includegraphics[width=\linewidth]{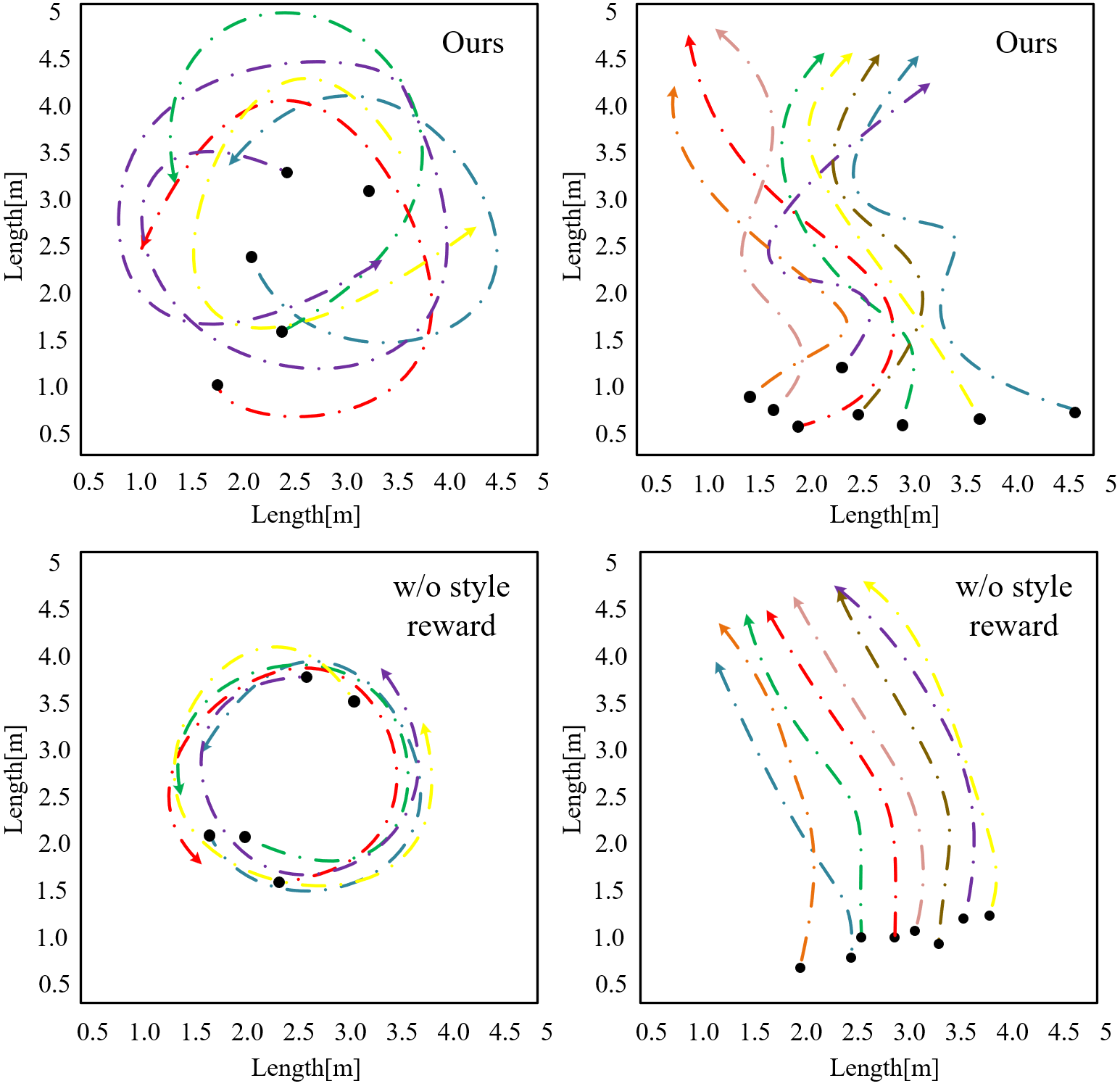}
                \vspace{-4mm}
        \caption{Visualization of fish trajectories produced by our method. The trajectories are generated by projecting the motion onto the XY plane. We sampled trajectories from five fish agents moving in a counterclockwise direction (\textbf{Left}) and eight fish agents moving in alignment (\textbf{Right}). A black dot indicates the initial position of each fish agent, while arrows depict their movement directions. Simulation time: 10 seconds.}
        \label{fig:traj_diversity}
    \end{minipage}
\end{figure}
In this section, we evaluate our method for schools of fish motion synthesis. Specifically, we compare our method with representative state-of-the-art (SOTA) methods including Boids~\cite{10.1145/37402.37406}, DeepFoids~\cite{NEURIPS2022_74fa9e6b}, and AMP~\cite{Peng_2021}. For a fair comparison, AMP uses the same video features for discrimination during adversarial imitation learning. More details about the settings of the compared methods are in Appendix D.5. We report metrics outlined in Sec~\ref{sub_sec:metrics} across four tasks: clockwise circling, counterclockwise circling, alignment, and aggregation, and compare our method with existing methods. A gallery showcasing the learned collective motions is shown in Fig.~\ref{fig:real2sim}. We also visualize several schools of fish with various movement patterns in Fig.~\ref{fig:Result_highlevel}. Readers are referred to the supplementary video for more details.

\paragraph{Cross-View Validation} 
Next, we employ cross-view validation, where reference videos offer multiple perspectives to verify the quality of generated motion sequences in the simulator. Specifically, we evaluate the Fréchet Inception Distance (FID) between unseen ground-truth views and generated videos from the simulator under the same unseen views. A lower FID score indicates closer generated motions to the reference distribution. FID is used for statistical analysis in Tab.~\ref{tab:FID} while qualitative analysis is shown in Fig.~\ref{fig:FID}. 
Tab.~\ref{tab:FID} illustrates that our approach exhibits remarkable performance in capturing the collective motion distributions compared to alternative methods.
As shown in Fig.~\ref{fig:FID}, our method exhibits robust generalization capability and consistent viewpoint preservation of our method when evaluated on unseen reference views. 

\begin{table}[ht]
  \caption{FID (lower is better) between the generated views and ground-truth views, with ground-truth views unseen during training.}
  \label{tab:FID}
  \footnotesize
 \vspace{-3mm}
\begin{tabular}{lcccc}
    \toprule
 Method & Clockwise & C-Clockwise & Aggregation  & Alignment \\
    \midrule
    Boids & $864.7$   &  $806.3$  &  $968.2$  &$891.5$ \\
    DeepFoids & $789.3$& $745.2$  & $928.9$ & $875.7$ \\
    AMP & $621.4$& $609.6$  & $685.4$ & $701.1$ \\
    Ours & $\textbf{534.5}$& $\textbf{501.9}$ & $\textbf{489.3}$  & $\textbf{523.3}$ \\
  \bottomrule
\end{tabular}
\end{table}

\paragraph{Skill Distribution.} For skill distribution, we report JS Divergence as the metric in Tab. \ref{tab:Distribution}. The results indicate that our method achieves lower JS Divergence values, suggesting that the learned distribution more closely aligns with the reference motion distribution, showing the effectiveness of the proposed CBIL framework.

\begin{table}[ht]
  \caption{Normalized JS Divergence~(lower is better) of various skill models across different tasks comparison. The clip length is $10$ frames.}
  \vspace{-3mm}
  \label{tab:Distribution}
  \footnotesize
\begin{tabular}{lcccc}
    \toprule
 Method & Clockwise & C-Clockwise & Aggregation  & Alignment\\
    \midrule
    Boids & $0.94$   &  $0.91$  &  $0.89$  &$0.96$ \\
    DeepFoids & $0.93$& $0.87$ & $0.78$  & $0.84$\\
    AMP & $0.79$& $0.69$  & $0.67$ & $0.79$ \\
    Ours & $\textbf{0.58}$& $\textbf{0.52}$  & $\textbf{0.42}$ & $\textbf{0.37}$ \\
  \bottomrule
\end{tabular}
\end{table}

\paragraph{Diversity Analysis.} To showcase the capacity of our method to acquire diverse collective movements from the reference, we utilize APD scores to assess the diversity of generated motion clips. Tab.~\ref{tab:Diversity} demonstrates that our method outperforms DeepFoids in terms of motion diversity. Notably, a higher APD score does not necessarily indicate better motion quality, as chaotic motions may still achieve high diversity scores. For instance, while AMP achieves a relatively higher APD score, it often produces chaotic motions. Fig.~\ref{fig:traj_diversity} shows the diverse trajectories produced by our method. All trajectories are produced from the same initial state. We generated $50$ trajectories, with each containing 50 frames for evaluation.

\begin{table}[t]
  \caption{Average Pairwise Distance (APD) scores (higher is better) of various skill models across different tasks comparison.}
    \vspace{-3mm}
  \label{tab:Diversity}
  \footnotesize
\begin{tabular}{lcccc}
    \toprule
 Method & Clockwise & C-Clockwise & Aggregation  & Alignment \\
    \midrule
    Boids & $32.8\pm1.79$   &  $41.5\pm1.63$  &  $23.1\pm1.92$  &$29.6\pm1.08$ \\
    DeepFoids & $45.7\pm0.93$ & $54.7\pm0.52$  & $34.2\pm0.69$ & $28.4\pm0.06$ \\
    AMP & $\textbf{127.9}\pm2.13$ & $\textbf{134.8}\pm1.14$  & $\textbf{142.7}\pm1.75$ & $\textbf{139.5}\pm1.10$ \\
    Ours & ${107.5}\pm1.22$& ${105.2}\pm1.18$ & ${119.6}\pm2.09$  & ${114.3}\pm1.62$ \\
  \bottomrule
\end{tabular}
\end{table}

\begin{figure}[htbp]
    \centering
    \begin{minipage}[t]{1\linewidth} 
        \centering
        \includegraphics[width=\linewidth]{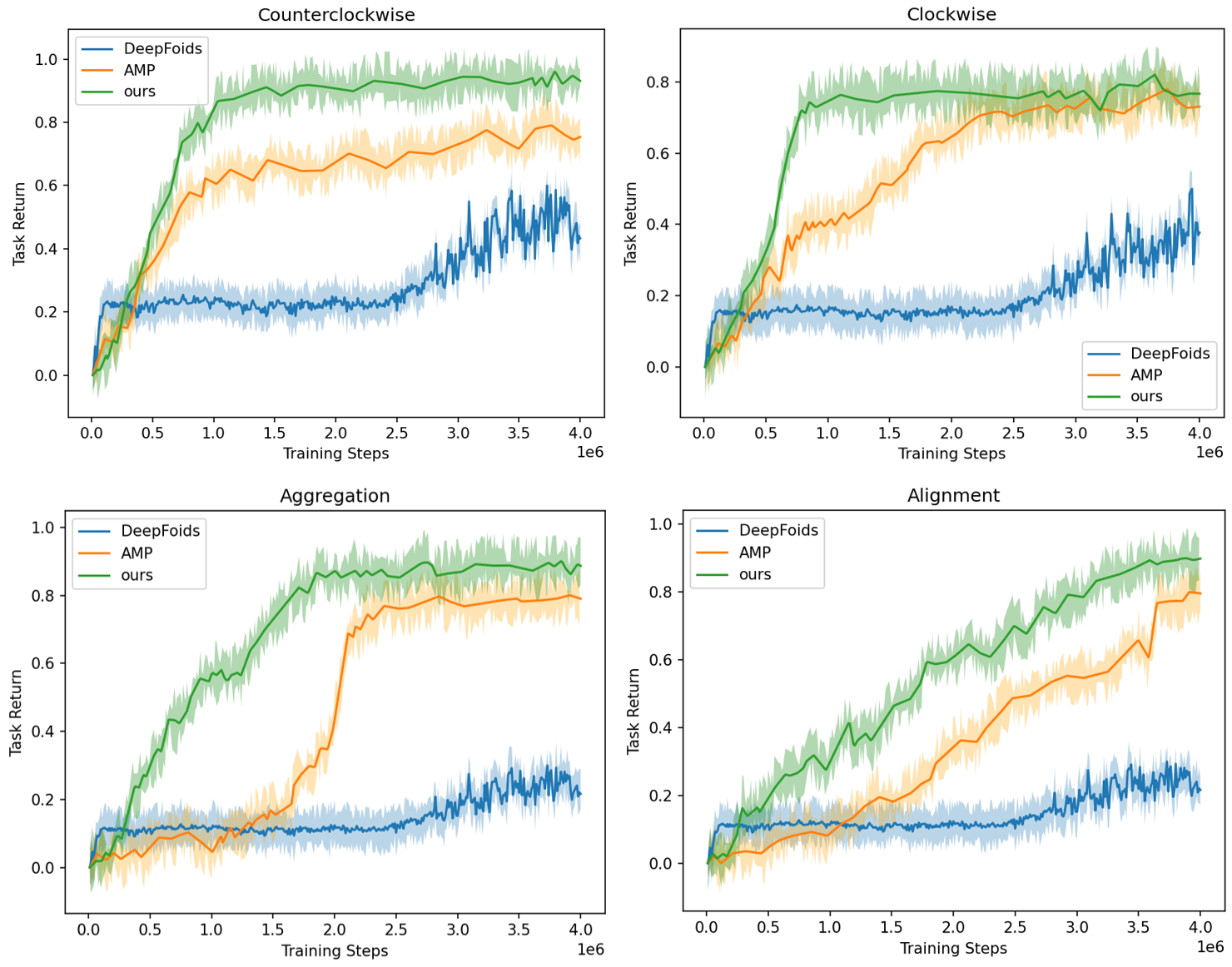}
                        \vspace{-4mm}
        \caption{Task return (higher is better) of different methods for different tasks.}
        \label{fig:Task Return}
    \end{minipage}
\end{figure}

\paragraph{Task Return.} We investigate the task return for the four animation tasks for learning-based approaches, i.e., DeepFoids, AMP, and ours. As summarized in Tab.~\ref{tab:Task Return} and Fig.~\ref{fig:Task Return}, existing methods often yield lower task returns and suffer from an unstable training process. Conversely, our method consistently surpass the previous methods in terms of task return and demonstrate better training stability and efficiency. Notably, task return alone does not comprehensively reflect method performance. For instance, although AMP shows similar task return performance in Fig.~\ref{fig:Task Return}, it generally performs worse across various metrics compared to our method. Furthermore, our method demonstrates improved training stability and efficiency compared with other methods.

\begin{table}[ht]
  \caption{Normalized task return (higher is better) of various skill models across tasks.}
    \vspace{-3mm}
  \label{tab:Task Return}
  \footnotesize
\begin{tabular}{lcccc}
    \toprule
 Method & Clockwise & C-Clockwise & Aggregation  & Alignment \\
    \midrule
    DeepFoids & $0.45\pm0.05$& $0.35\pm0.06$  & $0.32\pm0.02$ & $0.22\pm0.01$ \\
    AMP & $0.76\pm0.21$& $0.72\pm0.18$  & $0.85\pm0.15$ & $0.82\pm0.22$ \\
    Ours & $\textbf{0.78}\pm0.23$& $\textbf{0.92}\pm0.16$ & $\textbf{0.96}\pm0.14$  & $\textbf{0.93}\pm0.25$ \\
  \bottomrule
\end{tabular}
\end{table}

\section{More Collective Behavior Patterns}
\label{sec:applications}

\begin{figure}[t]
    \centering
    \begin{minipage}[t]{1\linewidth} 
        \centering
        \includegraphics[width=\linewidth]{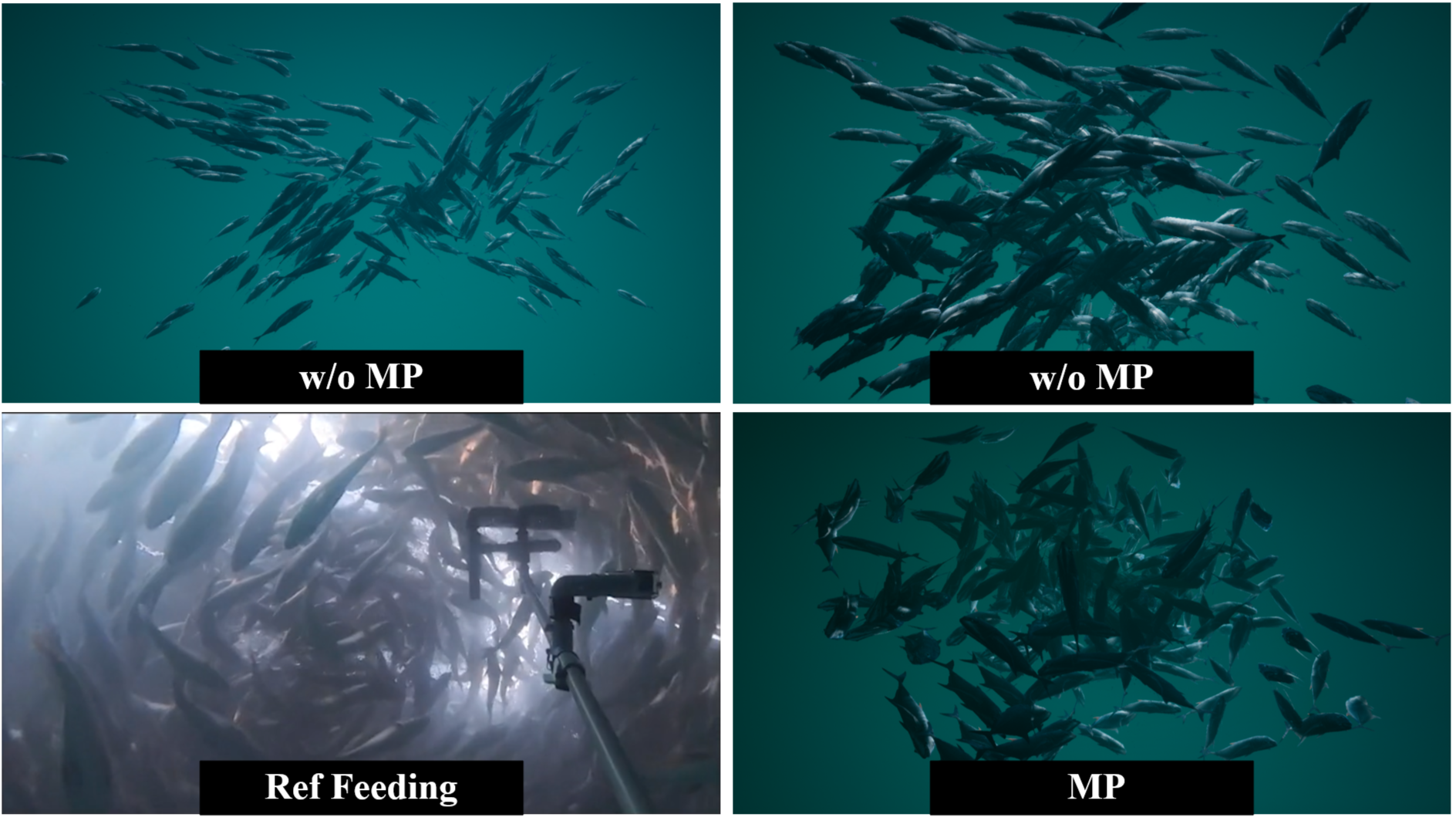}
                        \vspace{-4mm}
        \caption{Validation of Motion Prior Effectiveness in Fish Feeding Task. Our method (bottom row), trained with video-based motion priors, produces a more realistic animation of the school of fish compared to the pure rule-based approach trained without motion priors (top row). In the simulation, food is randomly generated and disappears once the closest fish agent remains within its sensor range for 3 seconds.}
        \label{fig:Feeding}
    \end{minipage}
\end{figure}

In this section, we present more animation results of fish schooling, including fish feeding and chasing. 
\paragraph{Fish Feeding}
Next, we validate the effectiveness of the motion prior learned from the videos for the fish feeding task by comparing our method with a counterpart that relies solely on rule-based feeding motion rewards without incorporating motion priors. As shown in Fig.~\ref{fig:Feeding}, when controlling a school of fish without motion priors, the fish exhibit only basic movement patterns, moving directly toward food attraction triggers (Top-left) and displaying minimal interaction with external stimuli even after converging (Top-right). These results lack realism, as the school appears unnaturally rigid, with less diverse behavior compared to the reference fish video captured in the fish farm (Bottom-left) where complex interactions with the environment and other fish agents are observed. In contrast, our method (Bottom-right), which learns the school's motion from videos, generates fish movements that naturally and realistically steer toward the food while maintaining diverse movement patterns. Additional animation results using our method with motion priors can be found in the supplementary video.

\begin{figure}[t]
    \centering
    \begin{minipage}[t]{1\linewidth} 
        \centering
        \includegraphics[width=\linewidth]{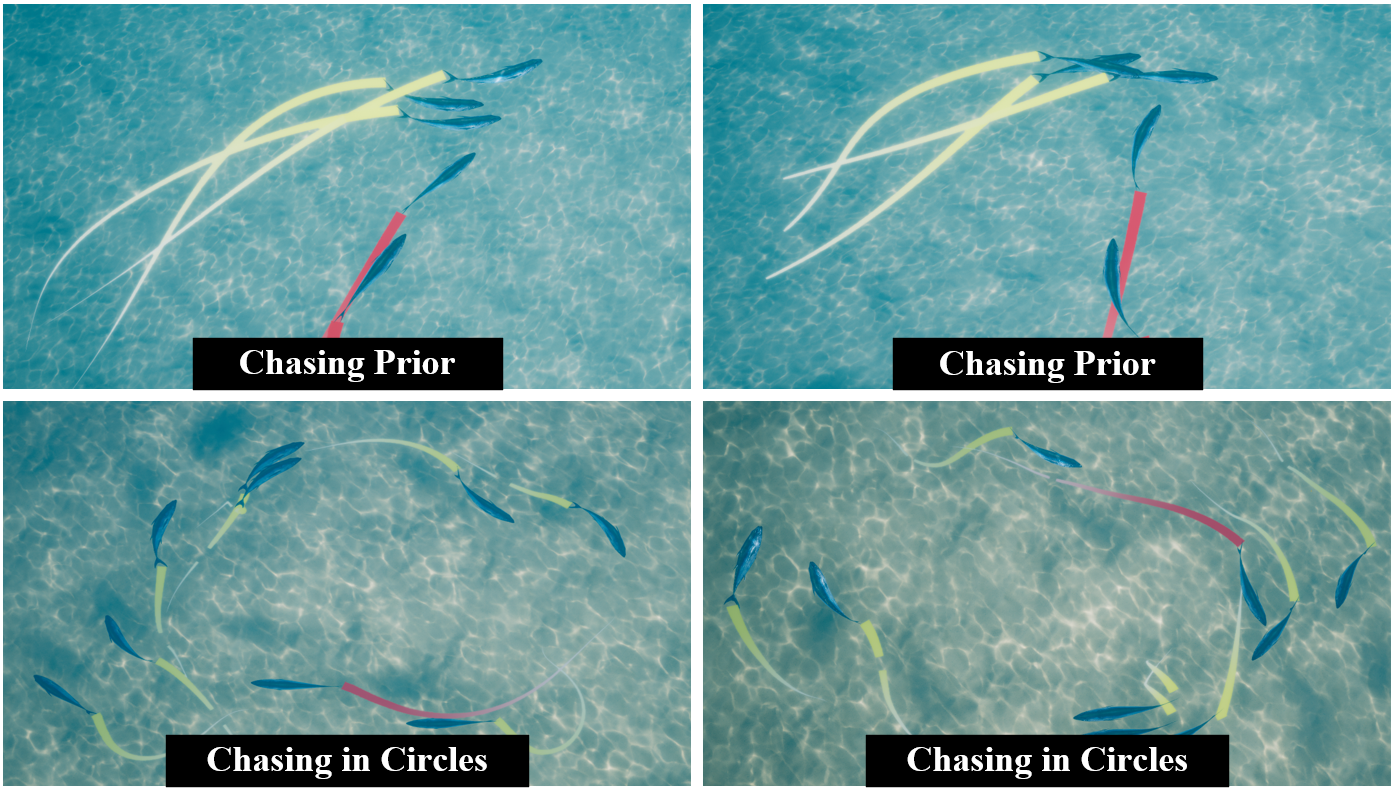}
                        \vspace{-4mm}
            \caption{Chasing in Circles. The school of fish transitions from chasing behavior to a consistent circling pattern, with the dominant fish (in red) chasing the subordinate fish (in yellow).}
        \label{fig:Chasing}
    \end{minipage}
\end{figure}

\paragraph{Chasing in Circles}
To further explore the general applicability of motion priors for various collective motion pattern syntheses, we present an animation result that employs a \textit{chasing} motion prior learned from videos to achieve a \textit{circling} task (Eq.~\ref{eq:circling}). In Fig.~\ref{fig:Chasing}, we first illustrate the learned chasing motion prior in the top row, where a school of fish exhibits chasing behavior, with dominant fish shown in red and subordinate fish in yellow. The bottom row demonstrates that, by applying the chasing motion prior along with a circling reward, the school of fish not only continues to exhibit the chasing behavior but also maintains a consistent circling pattern, validating the effectiveness of our learned motion prior. These reusable collective motion priors, learned from reference videos, enable more flexible control for achieving various animation tasks.

\section{Evaluation of Implicit State Clustering}
\label{sec:exp_ablation}

\paragraph{Ablation Study}
In the following, we investigate the impact of our implicit state clustering training strategy. As demonstrated in Tab.~\ref{tab:ablation_clustering}, the incorporation of implicit state clustering improves the performance across several key metrics, including FID, JS divergence, and task return. These improvements indicate better generative capabilities of our approach in achieving high-quality and diverse generative outcomes that are more aligned with real-world data.

\begin{table}[ht]
  \caption{The influence of implicit state clustering on fish school animation.}
    \vspace{-3mm}
  \label{tab:ablation_clustering}
  \footnotesize
\begin{tabular}{lccccc}
    \toprule
    Metrics                      & Settings & Clockwise & C-Clockwise & Aggregation & Alignment \\ 
    \midrule
    \multirow{2}{*}{FID}         &  w/o     &  578.2    &   551.6    &   529.4     &  565.9         \\  
                                 &  w/      &  534.5    &   501.9    &   489.3     &   523.3   \\ 
    \midrule
    \multirow{2}{*}{JS}          &  w/o     &   0.65    &    0.74    &    0.56     &   0.51       \\
                                 &  w/      &   0.58    &    0.52    &    0.42     &    0.37   \\ 
    \midrule
    \multirow{2}{*}{Task Return} &  w/o     &   0.72$\pm$0.14& 0.86$\pm$ 0.18  &  0.92$\pm$ 0.12          &     0.89$\pm$0.15      \\  
                                 &  w/      &  0.78$\pm$0.23 & 0.92$\pm$0.16 & 0.96$\pm$0.14 & 0.93$\pm$0.25 \\ 
    \bottomrule
\end{tabular}
\end{table}

\begin{figure}[htbp]
\vspace{-2mm}
    \centering
    \begin{minipage}[t]{1\linewidth} 
        \centering
        \includegraphics[width=\linewidth]{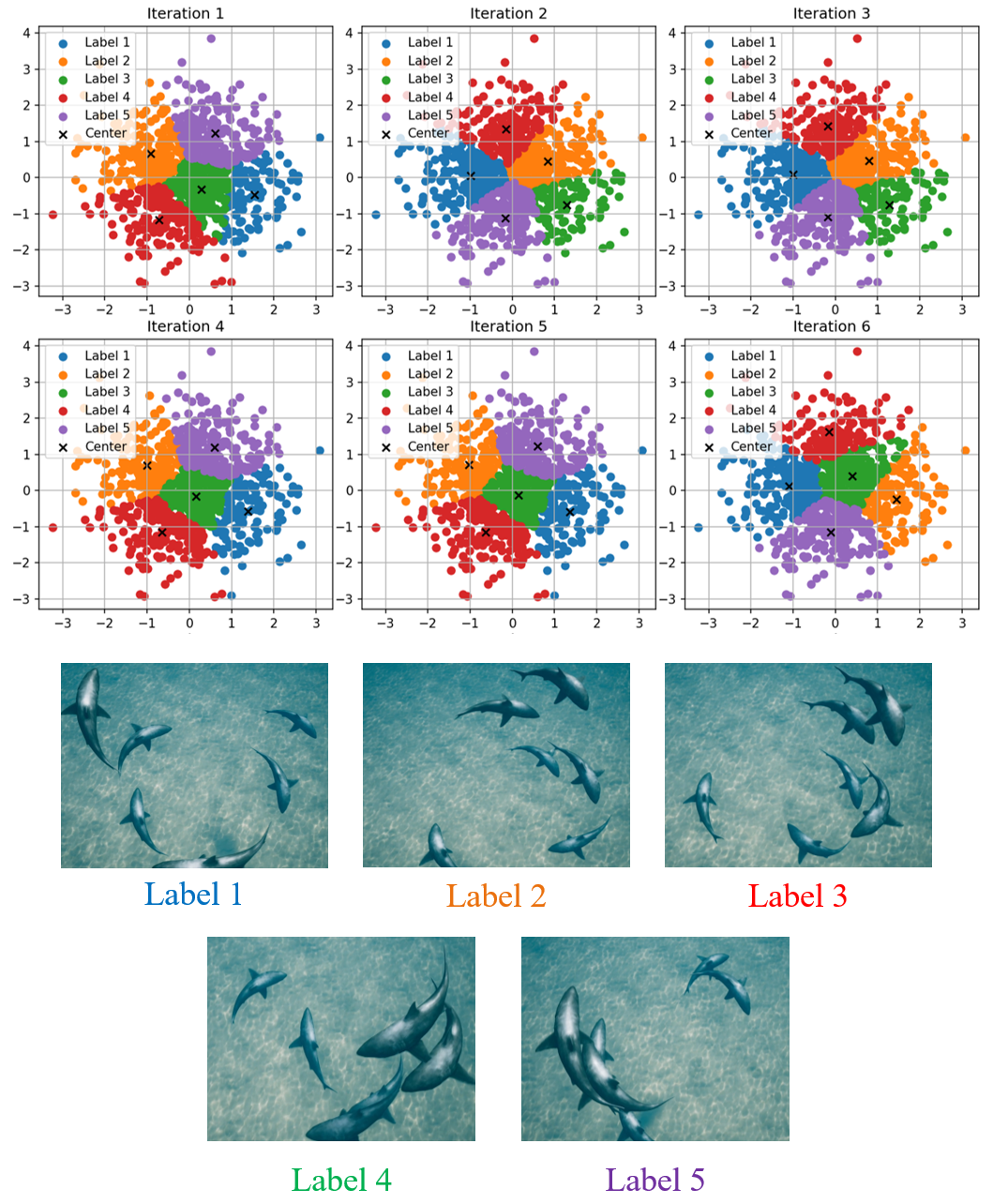}
        \vspace{-6.5mm}
       \caption{The visualization shows the implicit state clustering process for the task of clockwise circling. Here, iteration refers to the steps of the K-means algorithm to cluster the reference implicit states. After clustering, features of different movement patterns are grouped implicitly, with the corresponding collective motions visualized below.}
        \label{fig:Kmeans}
    \end{minipage}
\end{figure}

\paragraph{Feature Group Visualization}
\label{sec:feature_clustering}
In Fig.~\ref{fig:Kmeans}, we visualize the feature groups of different tasks during the implicit state clustering: during training, the implicit states are gradually clustered into different feature groups. Our training strategy effectively captures the features of each group during the training. We visualize the corresponding different motion patterns of a school of fish \textit{within} each feature group after clustering the latent features.

\section{Fish Abnormal Behavior Analysis}
\label{sec:abnormal}
The high-quality animation outputs with the simulated motion patterns can significantly enhance the motion analysis of fish schools. Detecting abnormal behavior~\cite{chong2017abnormal,li2019anomaly} is crucial in aquaculture, as sick fish can transmit contagious illnesses throughout the school, impacting the overall health of the population. Thus, detecting abnormal movement patterns in fish schools helps farmers mitigate losses from sick fish. However, in the expansive and ever-changing environment of marine ecosystems, the challenge of monitoring fish behavior to detect abnormalities presents significant challenges. Real-world data on fish behavior, particularly abnormal behavior indicative of environmental stress or disease, is not only scarce but also exceedingly difficult to capture due to the vastness and inaccessibility of aquatic ecosystems. This motivates us to train a detection model using synthetic data to simulate abnormal behaviors, supplemented with a small amount of expert-annotated real-world data.

\begin{figure}[!t]
        \vspace{-2mm}
    \centering
    \begin{minipage}[t]{1\linewidth} 
        \centering
        \includegraphics[width=\linewidth]{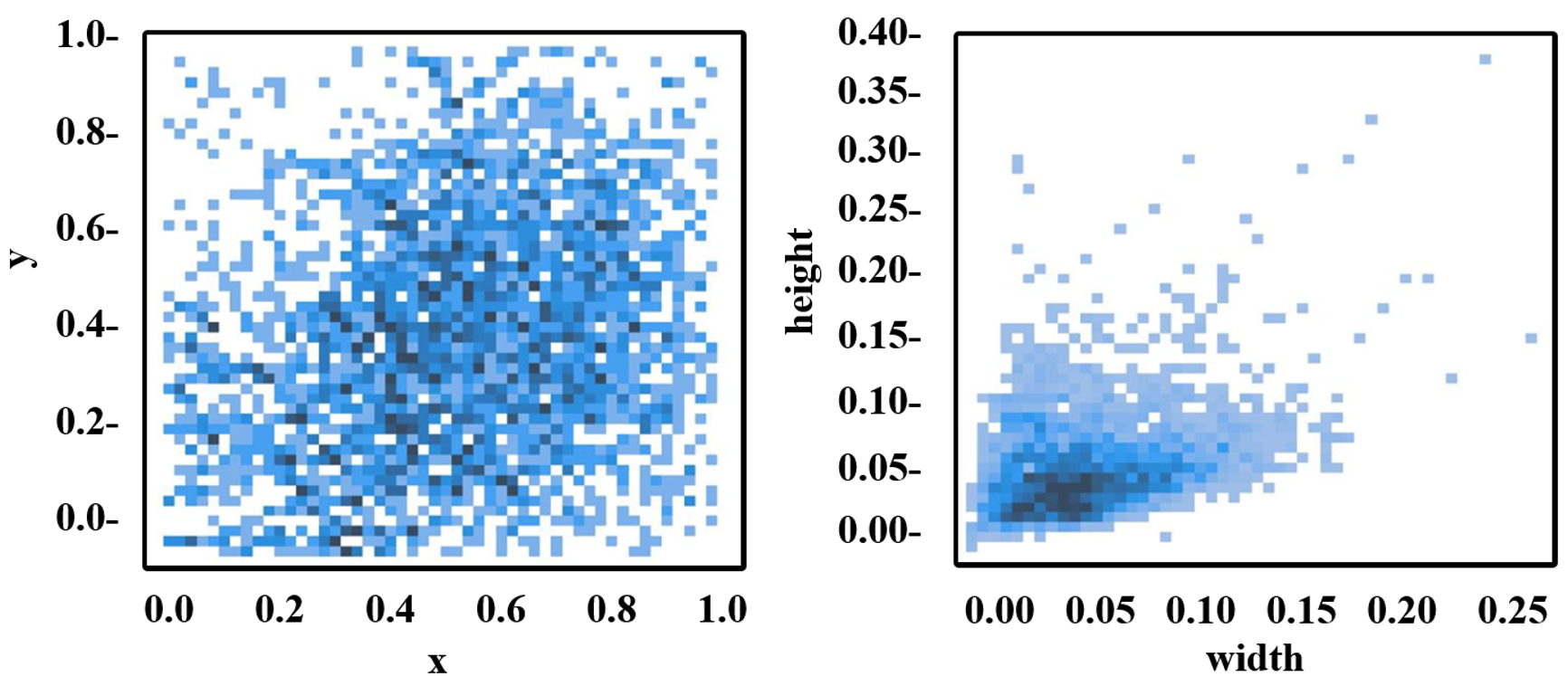}
        \vspace{-5mm}
        \caption{A total of $2901$ fish are annotated in our training data, comprising synthetic and real images. We present the distribution of the bounding box locations and sizes, normalized by image size.}
        \label{fig:training_data}
    \end{minipage}
\end{figure}

Specifically, we synthesized $60$ images using our system, in which some fish exhibit abnormal behavior. We randomly select 10\% of the fish from the group and place these selected fish $2$ meters away from the center of the group, while maintaining their previous velocity; this is because one of a typical abnormal behavior is to swim in isolation from the school.
Annotations are automatically produced through projection. Meanwhile, $40$ images were collected from our capture system at the fishing farm and manually annotated by fish farming experts. A total of $2901$ annotated bounding boxes were created. Of the total dataset of $100$ diverse images, $10\%$ were randomly selected and reserved as independent test samples to evaluate the accuracy of the detection process. The remaining $90$ images were used for subsequent data processing and network training. See Fig.~\ref{fig:training_data} for statistics on our training data. Building upon YOLOv8~\cite{chien2024yolov8am}, we train the model on our dataset with YOLOv8X pre-trained weights. The batch size is set to $16$ and the training epoch is set to $200$. For more details, please refer to Appendix E. 

\begin{figure}[t]
    \centering
    \begin{minipage}[t]{1\linewidth} 
        \centering
        \includegraphics[width=\linewidth]{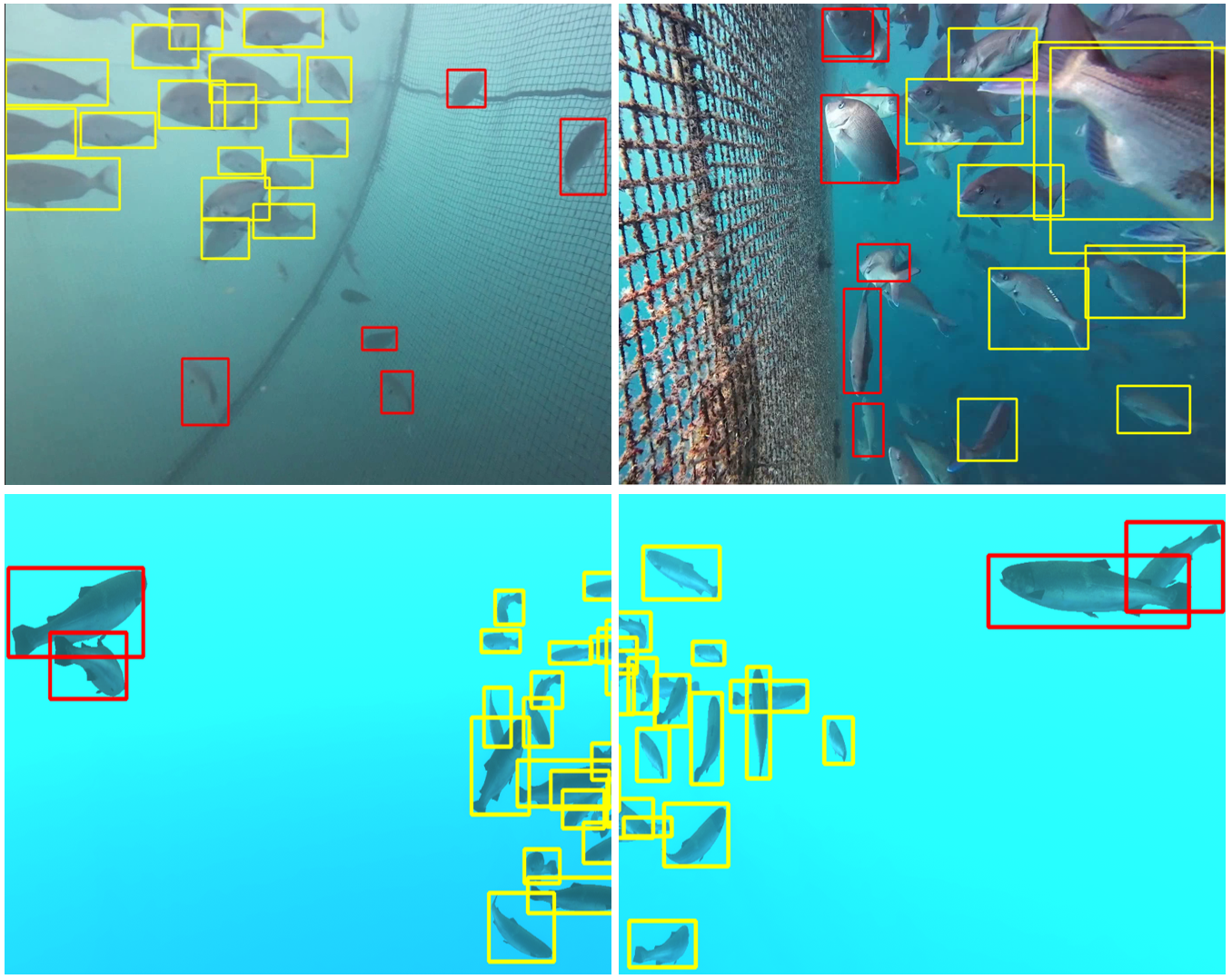}
        \vspace{-4mm}
        \caption{Samples of fish abnormal behavior detection from synthetic and real images. The \textcolor{yellow}{yellow} box denotes normal fish while the \textcolor{red}{red} box denotes detected abnormal behaviors.}
        \label{fig:sample_detection_result}
    \end{minipage}
\end{figure}
%


\paragraph{Evaluation.}
We report \textit{Precision}, \textit{Recall}, \textit{mAP@50}, and \textit{mAP@95} for the evaluation, following the methodology in~\cite{lin2014microsoft, redmon2016you, he2017mask}. We summarize the performance of the detection network using our data in Tab.~\ref{tab:detection}: the model trained with synthetic and real data achieves 
an overall precision of 96.8\% and a recall of 91\%. Fig.~\ref{fig:sample_detection_result} visualizes the detection results of the in-the-wild images.

\begin{table}[ht]
  \caption{Performance of fish abnormal behavior detection of our model on the test set.}
  \label{tab:detection}
  \footnotesize
  \vspace{-3mm}
\begin{tabular}{lcccccc}
    \toprule
  Class & Images & Instances & Precision & Recall & mAP50 & mAP50-95 \\
  \midrule
All     & 6 & 220 & 96.8\% & 91.0\% & 0.960 & 0.568 \\
Normal Fish & 6 & 201 & 95.0\% & 82.1\% & 0.925 & 0.580 \\
Abnormal Fish & 6 & 19 & 98.5\% & 100\% & 0.995 & 0.556 \\    
  \bottomrule
\end{tabular}
\end{table}

\begin{figure*}[t]
    \vspace{-1mm}
    \centering
    \includegraphics[width=\linewidth]{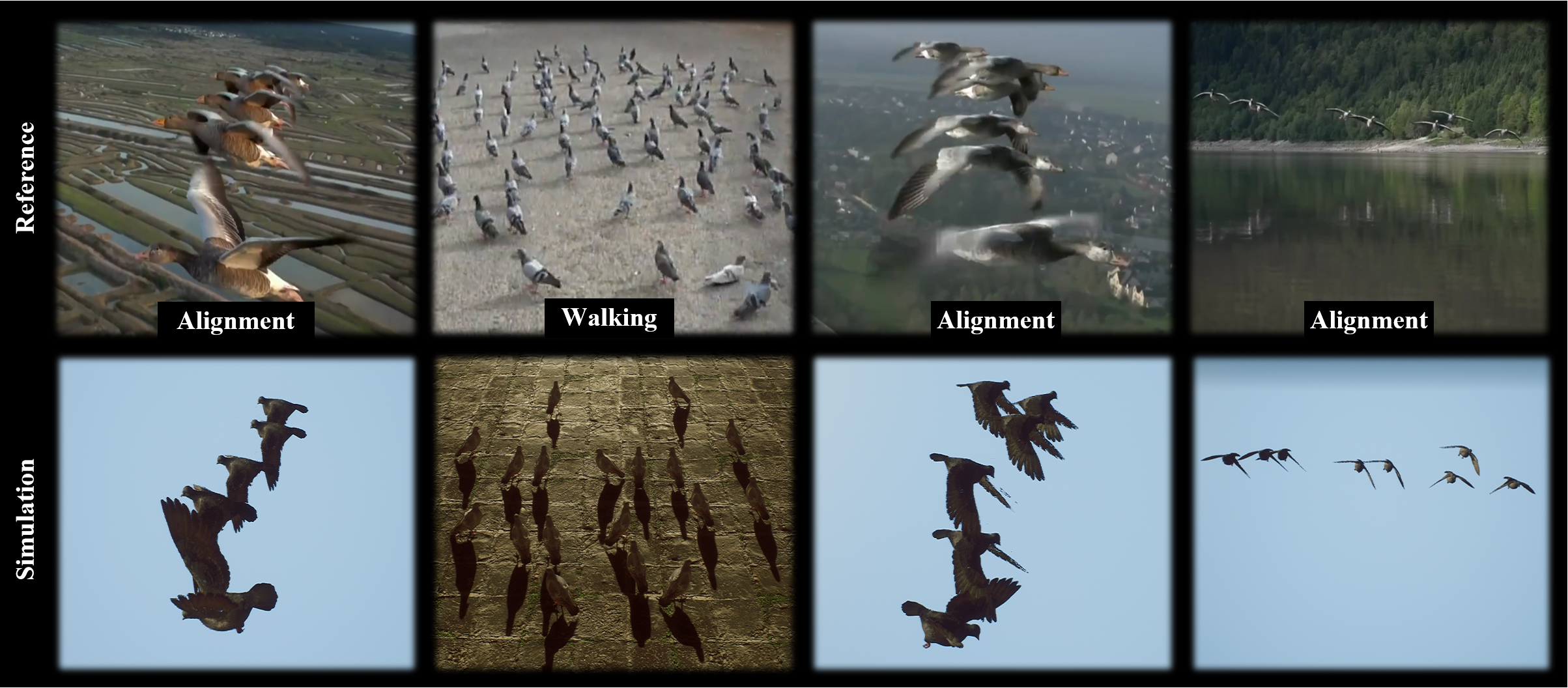}
    \vspace{-4.5mm}
    \caption{\name enables crowd animation across different species. Here, we demonstrate the effectiveness of our method in simulating a flock of birds, demonstrating that our framework reproduces a variety of behaviors learned from the reference videos.}
    \label{fig:bird}
    \vspace{-3mm}
\end{figure*}

\section{Generalization to Other Species}
In the following, we demonstrate the generalization capability of our method for simulating a flock of birds. We test with two scenarios: alignment and walking. For walking, we focus on 2D movement on the ground, with predefined specific patterns for smooth action transitions. For motions such as alignment during flying, the movement is controlled within 3D space. Each bird's trajectory is controlled by adjusting the rotation and speed of the agents using the predicted action, similar to the school of fish animation. For the detailed body movement, we make use of preset movements that match the root motion. As shown in Fig.~\ref{fig:bird}, the proposed framework generalizes well to the birds, producing realistic behavior learned from input videos. Additional animation results can be found in our supplementary video.

\section{Limitations and Future Works}
\label{sec:limitations}
Despite the advantages of \name in fish school animation described above, the system has limitations in the visual representation learning stage and the collective behavior imitation learning stage.  We first discuss such limitations and then the potential future works.      

\paragraph{Limitation with Visual Representation Learning.}
When the fish density becomes too dense for effective segmentation and analysis, our method may struggle to learn data-driven motion priors from 2D observations. Besides, our MVAE relies on comprehensive coverage of the latent variable distribution from reference videos, which necessitates the effort to generate diverse trajectories as discussed. If the MVAE fails to adequately cover the reference distribution, our method may struggle to achieve optimal performance.

\paragraph{Limitation with Adversarial Imitation Learning.}
\name operates in a GAIL style, which is still prone to mode collapse like other GAN-based methods, as revealed by~\cite{Peng_2021, Peng_2022, dou2023c}. Moreover, although \name could efficiently reproduce collective behavior from videos, the imitation learning for policy training remains sample-intensive. Data-efficient policy training methods~\cite{jena2021augmenting} could help improve learning efficiency. 

While \name has shown effectiveness in various tasks, it requires training different policies for different 
collective behaviors, e.g. circling;  the reference video clip that includes the specific collective behavior must be prepared and the policy needs to be trained using the corresponding loss function.  It would be ideal to develop a \textit{unified} model that is capable of handling fish with diverse species that allows smooth transitioning between different collective behaviors.      

%
\paragraph{Future Works} 
Our method primarily focuses on learning the general macroscopic trajectories of the school of fish; 
simulating the biomechanics of the fish and its interaction with the fluid using physical simulation could be beneficial for both computer animation and biology-related applications. 
For such purposes, 
techniques for simulating soft bodies and its interaction with fluid~\cite{newbolt2019flow, benchekroun2023fast, soliman2024going, verma2018efficient} could be useful.

Additionally, the proposed framework could be applied to replicating human crowd movements in videos. By combining the model with human pose estimation techniques, the accuracy of simulating individual body movements could be enhanced, which would provide a more detailed and realistic representation of crowd dynamics.

%

\section{Conclusion}
\label{sec:conclusion}
In this paper, we present Collective Behavior Imitation Learning (\name) for Fish, a scalable approach that directly learns fish school motions from videos, overcoming data sparsity and enhancing the effectiveness of imitation without relying on 3D motion trajectories. Our framework uses a Masked Video AutoEncoder (MVAE) to extract low-dimensional features in a self-supervised manner, enabling implicit state extraction from reference and simulated videos. During imitation learning, our framework effectively controls crowd movements and replicates the collective motion distributions from reference videos. By clustering implicit states and adaptively adjusting the discrimination reward weights based on group distributions, our imitation learning framework could robustly and efficiently capture diverse collective behaviors. The integration of bio-inspired rewards further provides regularization and stabilizes training with diverse reference data. As a result, \name produces various animations of fish schools. We further show its effectiveness across different species, such as birds, in crowd simulation. We also evaluate our system by synthesizing various fish animations for detecting abnormal fish behavior in in-the-wild videos.

\begin{acks}
This work is partly supported by the 
Research Grant Council of Hong Kong (Ref: 17210222),
the Innovation and Technology Commission of Hong Kong under the ITSP-Platform grants (Ref:  ITS/319/21FP, ITS/335/23FP) and the InnoHK initiative (TransGP project), and the contract research with Softbank Corp.
The authors are grateful to Sho Kakazu from SoftBank for creating the video, Michael Eastman from SoftBank for the narration recording, 
and Ryutaro Akasaka from Akasaka Fishery for supporting the field work on the sea. 
They also would like to thank SoftBank members for their experimental work on the sea.
\end{acks}
\bibliographystyle{ACM-Reference-Format}
\bibliography{ref.bib}
\end{document}